\documentclass[structabstract]{aa}  
\usepackage{graphicx}
\usepackage{txfonts}
\usepackage{natbib}
\usepackage[german,english]{babel}
\usepackage{psfig,lscape}
\usepackage{epsfig}
\usepackage{xspace}
\usepackage{epsf}
\newcommand{\ltapprox}{\raisebox{-0.5ex}{$\,\stackrel{<}{\scriptstyle\sim}\,$}}
\newcommand{\gtapprox}{\raisebox{-0.5ex}{$\,\stackrel{>}{\scriptstyle\sim}\,$}}
\newcommand{\zphot}{$z_{\rm phot}$~}

\newcommand\msol{${\rm M_{\odot}}$\xspace}

\newcommand\micron{$\mu$m}
\newcommand\msunyr{${\rm M_{\odot} yr^{-1}}$}
\def\aap{A\&A}

\def\apj{ApJ}
\def\apjl{ApJ}
\def\apjs{ApJS}
\def\mnras{MNRAS}

\begin{document}
   \title{Optical dropout galaxies lensed by the cluster A2667
\thanks{Based on observations collected at the European Southern Observatory,
Paranal, Chile, as part of the ESO 082.A-0163.
}}


   \author{N.~Laporte\inst{1,2}
          \and
R.~Pell\'o\inst{1,2} \and
D.~Schaerer\inst{3} \and
J.~Richard\inst{4} \and
E.~Egami\inst{5} \and
J. P.~Kneib\inst{6} \and
J. F.~Le Borgne\inst{1,2} \and
A.~Maizy\inst{1,2} \and
F.~Boone\inst{1,2} \and
P.~Hudelot\inst{7}  \and
Y.~Mellier\inst{7}
          }

   \institute{
Universit\'e de Toulouse; UPS-OMP; IRAP; Toulouse, France\\
\and
CNRS; IRAP; 14, avenue Edouard Belin, F-31400 Toulouse, France\\
\email{nicolas.laporte@ast.obs-mip.fr,roser@ast.obs-mip.fr}
\and
Geneva Observatory, 51, Ch. des Maillettes, CH-1290 Versoix, Switzerland\\
\email{Daniel.Schaerer@unige.ch}
\and
Institute for Computational Cosmology, Department of Physics, University of Durham, DH1 3LE, UK\\
\email{johan.richard@durham.ac.uk}
\and
Steward Observatory, University of Arizona, 933 North Cherry Avenue, Tucson,
AZ 85721, USA \\ 
\email{eegami@as.arizona.edu}
\and
Laboratoire d'Astrophysique de Marseille, CNRS - Universit\'e Aix-Marseille,
38 rue Fr\'ed\'eric Joliot-Curie, 13388 Marseille Cedex 13, France\\ 
\email{jean-paul.kneib@oamp.fr}
Institut d'Astrophysique de Paris, UMR7095 CNRS, Universit\'e Pierre \&  Marie
Curie, 98 bis boulevard Arago, 75014 Paris, France \\
\email{hudelot@iap.fr,mellier@iap.fr}
             }

   \date{Received ; accepted }

 
  \abstract
   {
{We investigate the nature and the physical properties of
ten $z$, $Y$ and $J-$dropout galaxies selected in the field of the lensing
cluster A2667.
}}
   {This cluster is part of our project aimed at obtaining deep
photometry at $\sim$0.8-2.5 microns with ESO/VLT HAWK-I and FORS2 on a
representative sample of lensing clusters extracted from our multi-wavelength
combined surveys with SPITZER, HST, and Herschel. The goal
is to identify a sample of redshift z$\sim$ 7-10 candidates accessible to
detailed spectroscopic studies.
}
   {
{The selection function is the usual dropout technique based on deep
$I$, $z$,$Y$, $J$, $H$ and $Ks$-band images (AB$\sim$26-27, 3$\sigma$),
targeting $z\gtrsim$7.5 galaxy candidates.
We also include IRAC data between 3.6 and 8 $\mu $m, and MIPS 24$\mu $m when
available.
In this paper we concentrate on the complete $Y$ and $J-$dropout sample among
the sources detected with a high S/N ratio in both $H$ and $Ks$ bands, as well
as the bright $z$-dropout sources fulfilling the color and magnitude selection
criteria adopted by Capak et al.\ (2011).
SED-fitting and photometric redshifts were used to further
constrain the nature and the properties of these candidates.
}}
   {10 photometric candidates are selected within the $\sim 7' \times 7'$
HAWK-I field of view ($\sim$ 33{\rm arcmin}$^2$ of effective area
once corrected for contamination and lensing dilution at $z\sim$7-10). All of them are
detected in $H$ and $Ks$ bands in addition to $J$ and/or IRAC
3.6$\mu$m/4.5$\mu$m images, with $H_{AB}$ ranging from 23.4 to 25.2, and have
modest magnification factors between 1.1 and 1.4.
Although best-fit photometric redshifts are obtained at high-$z$ for all
these candidates, the contamination by low-$z$ interlopers is expected to
range between $\sim$50-75\% based on previous studies, and on
comparison with the blank-field WIRCAM Ultra-Deep Survey (WUDS).
The same result is obtained when photometric
redshifts are computed using a luminosity prior, allowing us to remove half of
the original sample. Among the remaining galaxies, two additional sources
could be identified as low-$z$ interlopers based on a detection at 24$\mu $m
and on the HST $z_{850}$ band.
These low-$z$ interlopers are not well described by current spectral templates
given the large break, and cannot be easily identified based on
broad-band photometry in the optical and near-IR domains alone.
A good fit at z$\sim$1.7-3 is obtained at assuming a young stellar
population together with a strong extinction. Given the estimated dust
extinction and high SFRs, some of them could be also detected in the IR or
sub-mm bands.
}
   {After correction for contaminants,
the observed number counts at z$\gtrsim 7.5$ seem to be in agreement with
expectations for an evolving LF, and inconsistent with a constant LF since $z \sim 4$.
At least one and up to three candidates in this sample are expected to be
genuine high-$z$, although spectroscopy is still needed to
conclude.
}


   \maketitle
%
\section{Introduction}
\label{intro}

   Considerable advances have been made during the last years in the exploration
of the early Universe with the discovery of several z$\sim$6-7 galaxies
close to the end of reionization epoch (e.g. Kneib et~al. 2004, Stanway
et~al.\ 2004, Bouwens et~al.\ 2004, 2008, 2009, 2010, Iye et~al. 2006, Stark
et~al.\ 2007, Bradley et~al.\ 2008, Zheng et~al.\ 2009), whereas five-year
{\tt WMAP} results place the first building blocks at $z=11.0 \pm 1.4$,
suggesting that reionization was an extended process (Dunkley et~al.\ 2009). 
For now, very few galaxies beyond z$\sim$6.5 are spectroscopically confirmed
(Hu et~al.\ 2002, Cuby et~al.\ 2003, Kodaira et~al.\ 2003, Taniguchi
et~al.\ 2005, Iye et~al.\ 2006), and the samples beyond this limit are mainly
supported by photometric considerations (Kneib et~al.\ 2004, Bouwens
et~al.\ 2004, 2006, 2008, 2009, 2010; Richard et~al.\ 2006, 2008; Bradley et~al.\ 2008,
Zheng et~al.\ 2009, Castellano et~al.\ 2010, Capak et~al.\ 2011). Strong
evolution has been found in the abundance of 
galaxies between z$\sim$7-8 and z$\sim$3-4 (e.g. Bouwens et~al.\ 2008) based
on photometrically-selected samples, the star-formation rate (SFR) density
being much smaller at very high-z up to the limits of the present surveys, in
particular towards the bright end of the Luminosity Function (LF). A similar
evolutionary trend is observed in narrow-band surveys searching for
Lyman-alpha emitters (e.g. Iye et~al.\ 2006,
Cuby et~al.\ 2007, Hibon et~al.\ 2010).  

   Lensing clusters of galaxies provide a unique and priviledged view of the
high-redshift Universe. This technique, often referred to as ``gravitational
telescope'', was first proposed by Zwicky (1937). It has proven highly
successful in identifying several of the most distant galaxies known
today thanks to magnifications by typically 1-3 magnitudes (e.g. Ellis et
al.\ 2001, Hu et al.\ 2002, Kneib et al.\ 2004, Bradley et al.\ 2008, Zheng
et~al.\ 2009).  

   Our project is based on the photometric pre-selection of high-z candidates
in lensing clusters using the Lyman-break technique (LBGs, e.g. Steidel
et~al.\ 1995), which has been proved successful to identify star-forming
objects up to z$\sim$ 6 (Bunker et~al.\ 2004, Bouwens et al.\ 2004 to 2009),
as well as photometric redshifts. The long term goal is to 
substantially increase the present sample of redshift z$\sim$ 7-12 galaxies,
and to study their physical properties and the star-formation activity using a
multi-wavelength approach. 

   This paper presents the first results of the short-term ongoing
survey aimed at completing a deep photometric survey at $\sim$0.8-2.5 microns
with HAWK-I at ESO/VLT on a representative sample of strong-lensing clusters at
intermediate redshift, extracted from our multi-wavelength combined survey
with SPITZER/IRAC+MIPS, HST (ACS/WFC3/NIC3), Herschel OT Key Program, and
sub-mm coverage.  
Herschel data on this field have been recently obtained with PACS and SPIRE as
part of the Herschel Lensing Survey (HLS, PI. E. Egami; see also Egami et
al.\ 2010). These results and data will be described in a forthcoming paper. 
The presence of a strong
lensing cluster within the large field-of-view of HAWK-I ($\sim 7'\times7'$) 
is expected to optimize the global efficiency of the survey by
combining in a single shot the benefit of gravitational magnification and a
large effective surveyed volume (see also Maizy et~al.\ 2010). 

   In this paper we concentrate on the complete sample of $Y$ and $J-$dropout 
sources selected in the field of the lensing cluster A2667 
(Abell 2667, $\alpha$=23:51:39.35 $\delta$=$-$26:05:03.1 J2000, $z=0.23$). 
For comparison we also examine the bright $z$-dropout sources fulfilling the color and
magnitude selection criteria adopted by Capak et al.\ (2011).
All the candidates are selected to represent star-forming
galaxies at $z\gtrsim 7.5$, and to be bright enough for spectroscopic follow
up, although a large fraction of galaxies selected in this way could 
be low-$z$ contaminants.
SED-fitting and photometric redshifts are used to further constrain the nature
and the properties of these candidates. The results 
achieved on the luminosity functions in the $z\gtrsim 7.0$ redshift domain will
be presented elsewhere. 

   In Sect.\ \ref{data} we summarize the photometric observations and data
reduction. We also describe the construction and analysis of photometric
catalogs. The selection of different dropout candidates is presented in
  Sect.\ \ref{selection}. 
The properties of the
selected candidates, including spectral energy distributions (hereafter SEDs),
SED-fitting results and photometric redshifts are detailed in
Sect.\ \ref{results},
together with a discussion on the reliability of the different high-$z$
candidates when photometric redshifts include a luminosity prior.
In Sect.\ \ref{discussion} we discuss the global properties of
this sample, 
in particular expected versus observed number counts. We compare our
results with previous findings and we discuss on the nature of low-$z$
contaminants.
Conclusions are given in Sect.\ \ref{conclusions}.
The concordance cosmology is adopted throughout this paper, with
$\Omega_{\Lambda}=0.7$, $\Omega_{m}=0.3$ and $H_{0}=70\ km\ s^{-1}\ Mpc^{-1}$.
All magnitudes are quoted in the AB system (Oke \& Gunn 1983). Table
~\ref{tab_phot} presents the conversion values between Vega and AB systems for
our photometric dataset. 

\section{Observations and data analysis}
\label{data}

\subsection{Ground-based optical and near-IR imaging}

  The selection of high-z candidates is based on deep optical and near-IR
imaging. A2667 was observed with HAWK-I in the near-infrared domain ($\sim$ 0.9 to
2.2 $\mu$m, covering the 4 bands $Y$, $J$, $H$, and $Ks$), and with FORS2 in
$I$ and $z$ bands, between October and November 2008. 
Data reduction and processing included photometric calibration, bias
subtraction, flat-fielding, sky subtraction, registration and combination of
images using {\sc IRAF}, closely following the general procedures described in
Richard et al.\ (2006). 
Table~\ref{tab_phot} summarizes the properties of the photometric dataset used
in this paper. 

For the FORS2 data, we used a standard flat-field
correction and combination of the individual frames with bad-pixel rejection.
In addition to the $z$ band image matching the HAWK-I field (hereafter
$z_1$), an older image of similar quality centered on the cD galaxy, obtained
in June 2003 (71.A-0428, hereafter $z_2$) was also used to confirm the
non-detection of high-$z$ candidates in the common area. 

For HAWK-I data, we used the ESO pipeline
\footnote{See {\tt http://www.eso.org/sci/data-processing/software/pipelines/index.html}}
to process and combine all individual images, refining the offsets
between the different epochs of observations when producing the final
stack. This procedure performs a 2-step sky subtraction, using masks to 
reject pixels located on bright objects, similar to the XDIMSUM package 
as described in Richard et al. (2006). The photometric calibration was checked
against 2MASS stars present in the field of view, taking into account the
relative flat-field normalisations between each one of the 4 HAWKI chips.
Before combining, we applied individual weight values according to:
$weight \propto (ZP\times var\times {s}^2)^{-1}$
where individual zero-point $ZP$ and seeing $s$ values were obtained from 
the brightest unsaturated stars around the field, and pixel-to-pixel variance
$var$ was derived through statistics within a small region free of objects. 

    Photometric zero-points were derived from LCO/Palomar
NICMOS standard stars (Persson et al.\ 1998). The final accuracy of our 
photometric calibration has been checked by comparing the observed colors of
cluster elliptical galaxies, measured on images matched to the worst seeing in
our data (i.e. $\sim0.9$\arcsec in the $z_1$ band), to match expectations
based on the empirical SED template compiled by Coleman, Wu and Weedman
(CWW, 1980). According to this check, we expect our final photometric catalog to be
accurate to about 0.05 mags throughout the wavelength domain between $I$ and $Ks$.

   All HAWK-I images were registered and matched to a common seeing using a
simple gaussian convolution in order to derive magnitudes and colors, the
worst case being the $J$ band image. Astrometric calibration was performed in
a standard way (see e.g. Richard et al.\ 2006), reaching an absolute accuracy
of $\sim 0.2$ \arcsec\ for a whole HAWK-I field of view.
Images in $I$ and $z$ bands were 
registered to the HAWK-I images using standard {\sc IRAF} procedures for rotation,
magnification and resampling of the data. These images were mainly used to
exclude well-detected low-$z$ interlopers. 

   All high-$z$ candidates are expected to be detected in the $H$ and
$Ks$ bands and to be undetected in $I$ and $z$ bands. Therefore, the original
$H$ and $Ks$ band images were combined together to create an $H+Ks$ detection image
with excellent seeing quality ($\sim0.47$\arcsec). Also the original
unregistered images in $I$ and $z$ bands were used for the visual inspection
of the Y-dropout candidates. Error bars and non-detection criteria were also
determined on the original images (see below). Photometry for the $z$ and
$Y$-dropout sample was also obtained with near-IR images matched to the worst
seeing in our data in order to check for consistency of observed colors. 

   We used the $SExtractor$ package version 2.8 (Bertin \& Arnouts 1996) for
detection and photometry. Magnitudes were measured in all images with the
$SExtractor$ ``double-image'' mode, using the $H+Ks$ detection image as a
reference. Magnitudes were measured within different apertures ranging from 1
to 2 \arcsec diameter, as well as MAG\_AUTO magnitudes. We checked that colors
derived with different choices of aperture and magnitude types are consistent
with each other within the errors. Photometric errors were measured using the
typical RMS in the pixel distribution of the original images (without any
seeing matching or rescaling), within apertures of the same physical size as
for flux measurements (either aperture or MAG\_AUTO magnitudes). Photometric
errors measured on the original images were also used to compute $3\sigma$
limiting magnitudes in each band, reported in Table~\ref{tab_phot}. Errors in
colors were derived from the corresponding magnitude errors added
quadratically. 

\begin{figure}
   \centering
  \includegraphics[angle=270,width=0.50\textwidth]{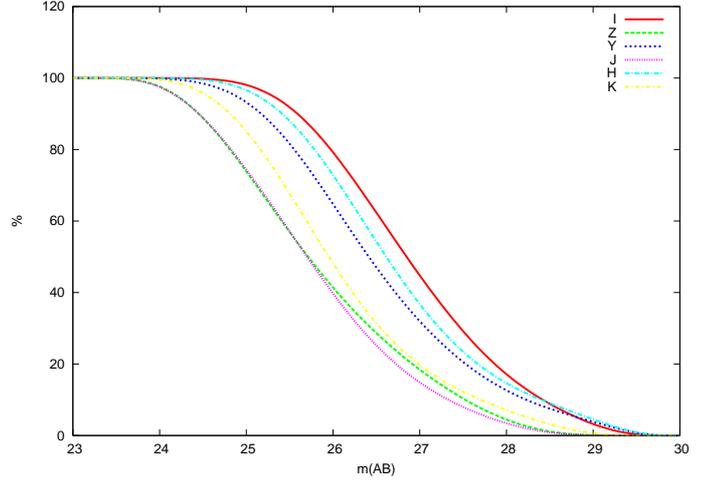}
      \caption{Completeness levels obtained from simulations for 
the different bands used for high-z sample selection: 
$I$ (solid red line), $z$ (dashed green line),
$Y$ (dotted blue line), $J$ (dotted magenta line),$H$ (dot-dashed cyan line),
and $Ks$ (dot-dashed yellow line).
              }
         \label{completeness}
   \end{figure}

    Completeness values for point-sources detected in the different bands were
obtained through Monte-Carlo techniques. Artificial stars (i.e., seeing limited sources)
were added at $\sim$700 different random locations on our images, and then extracted
using the same parameters for detection and photometry as for astronomical
sources. The seeing was measured on the original co-added images.
The corresponding results are shown in Fig.~\ref{completeness} and
Table~\ref{tab_phot}.   

\subsection{Other imaging observations}
\label{spitzer}

   In addition to the images used for high-z sample selection, SED analysis is
also supported by additional data when available for the candidates, namely:

\begin{itemize} 

\item Images of Abell 2667 were obtained 
in the 3.6, 4.5, 5.8 and 8.0$\mu $m bands
as part of the GTO Lensing
Cluster Survey (Program 83, PI. G. Rieke), using the Infrared Array Camera
(IRAC; Fazio et al.\ 2004) onboard the Spitzer Space Telescope (Werner et
al.\ 2004). In addition, deeper exposures were obtained at 3.6 and 4.5$\mu $m
in August 2009 (Program 60034, PI. E. Egami). 
All these images were processed according to Egami et al.\ (2006). IRAC magnitudes
were measured within a 2\arcsec \ diameter aperture and corrected according to
the Instrument Handbook (v.1.0, February 2010). The field of view is 5.2 \arcmin
$\times$ 5.2 \arcmin. 

\item We also gathered 24$\mu$m images in this field obtained with MIPS (Rieke et
al.\ 2004), processed as described by Rigby et al.\ (2008). 

\item A deep HST F850LP/ACS image is also available on the cluster core (8.8 ksec,
PI. R. Ellis; see also Richard et al.\ 2008). Photometry in this band is
refered as $z_{850}$ hereafter. 

\end{itemize} 

The properties of the complete photometric dataset used in this paper are
summarized in Table~\ref{tab_phot}. 3.6 and 4.5$\mu $m images refer to the
most recent and better quality dataset. 
%

\begin{table}
\caption{\label{tab_phot}Photometric dataset: ground-based optical and near-IR
  imaging used in the selection of high-z candidates (top) and space-based complementary
  data (bottom)}
\centering
\begin{tabular}{lrrrcccc}
\hline
Filter &  $\lambda_{eff}$& $C_{AB}$ & $t_{exp}$ & pix 
& m(3$\sigma$)& m(50\%) & seeing \\
       & [nm]           & [mag]    & [ksec]  & [\arcsec] & [mag]&
[mag]  & [\arcsec] \\
\hline
$I$    & 793 & 0.45 & 13.0 & 0.126 & 27.5 & 26.8 & 0.47 \\
$z_1$  & 920 & 0.54 & 12.7 & 0.126 & 26.1 & 25.7 & 0.91 \\
$z_2$  & 920 & 0.54 & 13.2 & 0.126 & 26.0 & 25.7 & 0.54  \\
$Y$   & 1021 & 0.62 &  8.6 & 0.106 & 26.9 & 26.3 & 0.61  \\
$J$   & 1260 & 0.95 &  9.2 & 0.106 & 26.3 & 25.7 & 0.64  \\
$H$   & 1625 & 1.38 & 25.3 & 0.106 & 26.8 & 26.5 & 0.46  \\
$Ks$  & 2152 & 1.86 & 11.0 & 0.106 & 25.9 & 25.8 & 0.47  \\
\hline
    &  &  &  &  &  & Ref. &  \\
\hline
$z_{850}$ & 9106 & 0.54 &  8.8 & 0.05& 26.1& A&   \\
3.6$\mu$m & 3575 & 2.79 & 16.8 & 1.2 & 25.1 & &     \\
4.5$\mu$m & 4528 & 3.25 & 17.4 & 1.2 & 25.2 & &    \\
5.8$\mu$m & 5693 & 3.70 & 2.4  & 1.2 & 22.7 & B&    \\
8.0$\mu$m & 7958 & 4.37 & 2.4 & 1.2 & 22.8 & B &   \\
24$\mu$m  &23843 & 6.69 & 2.7 & 2.55& 18.7 & C &   \\
\hline
\end{tabular}
\tablefoot{Information given in this table: filter identification, filter effective wavelength, and AB
correction ($m_{AB}=m_{Vega}+C_{AB}$), total exposure time, 
pixel size, 3 $\sigma$ limiting magnitude and average seeing. For the filters used in the
high-z sample selection, the 50\% completeness level is also indicated (point
source, 1.3 \arcsec \ diameter aperture). References: 
A. Richard et al.\ (2008),
B. Egami et al.\ (2004),
C. Rigby et al.\ (2008)
}
\end{table}


\begin{table*}[h]
\caption{\label{tab_catalog1}Catalogue of $z$, $Y$ and $J$-dropouts in A2667.}
\centering
\begin{tabular}{rccccccccccccrcc}
\hline\hline
Source &RA &Dec  & $H$ & $\Delta H$ & Stell. & FWHM & $\mu$ & $\mu$ &
$\Delta\mu$ & Q1 & Q2 & Q3 & Q & $\chi^2_{opt}$ & Final \\
       &(J2000)& (J2000) &  &  &  & [\arcsec]& high-$z$   & low-$z$ &  &  &
 &  &  &  & \\
\hline
z1 & 23:51:45.837& -26:7:07.20 & 23.59 & 0.03 &  0.03 & 0.74& 1.169 & 1.119 &0.018& 3 & 3 & 1 & 7-I  & -0.03 & (low) \\ 
Y1 & 23:52:00.157& -26:8:30.31 & 23.35 & 0.03 & 0.03  & 0.95& 1.028 & 1.012 &0.003& 1 & 1 & 2 & 4-III & 1.00 & low \\ 
Y2 & 23:51:57.156& -26:4:02.37 & 23.45 & 0.03 & 0.40 & 0.58 & 1.108 & 1.031 &0.007& 2 & 2 & 1 & 5-III & 1.19 & low \\
Y3 & 23:51:43.332& -26:8:00.89 & 23.68 & 0.04 & 0.02 & 1.24 & 1.128 & 1.059 &0.012& 3 & 3 & 2 & 8-I & 0.54 & high \\
Y4 & 23:51:35.750& -26:7:10.65 & 23.80 & 0.04 & 0.12 & 0.74 & 1.371 & 1.111 &0.020& 1 & 3 & 2 & 6-II & 0.64 & high \\
Y5 & 23:51:54.448& -26:3:13.65 & 23.98 & 0.04 & 0.03 & 0.96 & 1.149 & 1.037 &0.007& 2 & 2 & 2 & 6-II & 0.68 & ? \\
Y6 & 23:51:53.437& -26:4:29.94 & 24.37 & 0.08 & 0.02 & 1.81 & 1.167 & 1.067 &0.011& 1 & 1 & 3 & 5-III & 0.07 & low  \\
Y7 & 23:51:56.568& -26:7:51.45 & 24.27 & 0.05 & 0.02 & 1.13 & 1.045 & 1.021 &0.005& 3 & 3 & 2 & 8-I & 0.00 & low \\
Y8 & 23:51:37.151& -26:2:30.46 & 25.40 & 0.08 & 0.97 & 0.46 & 1.183 & 1.140 &0.018& 1 & 2 & 3 & 6-II & 0.58 & low  \\
J1 & 23:51:34.855& -26:3:32.74 & 25.21 & 0.08 & 0.97 & 0.55 & 1.299 & 1.398 &0.033& 3 & 3 & 2 & 8-I & 0.60 & (low) \\
\hline
\end{tabular}
\tablefoot{Information given in this table: (1) Object Identification; (2,3)
  $\alpha$ and $\delta$ J2000; (4,5) AUTO magnitudes and error bars in $H$ band;
  (6,7) SExtractor stellarity index and maximum FWHM measured on the detection
  $H+Ks$ image;
(8, 9, 10) lensing magnification (high-$z$ and low-$z$ hypothesis) and
associated uncertainty; 
(11, 12, 13) quality grade for high-$z$ candidates according to 3
criteria: possible contamination by close neighbours (Q1), completeness and quality of
photometric SED (Q2), and UV luminosity (Q3); 
(14) total grade Q; (15) optical $\chi^2$, and (16) tentative classification 
between low and high-$z$ using a stringent prior in magnitude. In the case of
z1 and J1 (in brackets), the low-$z$ indentification is forced based on the
detections in the 24$\mu$m and $z_{850}$ bands respectively for z1 and J1
(see Sect.~\ref{properties}).
}
\end{table*}

\begin{table*}[h]
\caption{\label{tab_catalog2}Photometric catalogue of bright $z$, $Y$ and $J$-dropouts in A2667.}
\centering
\begin{tabular}{lccccccccccc}
\hline\hline
Source & $I$ & $z$ & $Y$ & $J$ & $H$ & $Ks$ & 3.6$\mu$m
& 4.5$\mu$m & 5.8$\mu$m &  8.0$\mu$m & 24$\mu$m \\
   &  &  &  &  &  &  & 
&  &  &   &  \\
\hline
z1(*) & $>$28.7 & $>$27.2 &  25.63    &  23.29 &  23.59   & 23.00 & 21.79  & 21.38 & - &  21.41& 17.57 \\ 
      &         &         & $\pm$0.15 & $\pm$0.03&$\pm$0.03 &$\pm$0.03 &$\pm$0.01 &$\pm$0.01 &  & $\pm$0.10 & $\pm$0.12 \\
Y1 & $>$28.7 & -       &  27.04    & 24.08     & 23.35    &  22.73& -  & - & -  &  -  & -  \\ 
   &         &         & $\pm$0.35 & $\pm$0.07 &$\pm$0.03 &$\pm$0.03 &   &   &  &   & \\
Y2 & $>$28.7 & $>$27.2 & $>$28.1   &24.17      & 23.45    &22.81 &  -  &21.80 & -  &  -  & -  \\
   &         &         &           &$\pm$0.15  &$\pm$0.03 &$\pm$0.03 &  &$\pm$0.01 &    &   & \\
Y3(*) & $>$28.7 & $>$27.2 & 25.51     &23.85 & 23.68    & 23.07& 22.84& 22.52 & - & -  & $>$19.93 \\
   &         &         & $\pm$0.18 &$\pm$0.07  &$\pm$0.04 &$\pm$0.04  &$\pm$0.04  &$\pm$0.03&  &   & \\
Y4(*) & $>$28.7 & 26.42 &$>$28.1    & 24.59 & 23.80    &23.55 &23.19 &22.91 & $>$23.93 & $>$23.98  & $>$19.93 \\
   &         &$\pm$0.50&           &$\pm$0.11  &$\pm$0.04 &$\pm$0.05  &$\pm$0.06  &$\pm$0.04&  &   &   \\
Y5(*) & $>$28.7 & $>$27.2& $>$28.1   &  24.31    & 23.98    & 23.28 & 22.74 & 22.47 &  -  &  -  & -  \\
   &         &         &           &$\pm$0.08  &$\pm$0.04 &$\pm$0.04  &$\pm$0.04  &$\pm$0.03&  &   & \\
Y6(*) & $>$28.7 &$>$27.2 &26.91      & 24.84     & 24.37    &23.75 &  -  & - &  -  &  -  & -  \\
   &         &         &$\pm$0.33  &$\pm$0.13  &$\pm$0.08 &$\pm$0.06&  &   &  &  &    \\
Y7 & $>$28.7 & $>$27.2 & $>$28.1   &25.22      &24.27     &23.75&   22.53 & 22.68 & - & -  & 18.78 \\
   &         &         &           &$\pm$0.12  &$\pm$0.05 &$\pm$0.08 & $\pm$0.03 &$\pm$0.03&  &   & $\pm$0.36\\
Y8(*) & $>$28.7 & $>$27.2 & 26.86     & 25.58     &25.40     & 25.08&   -  & -   &21.24 &22.12& $>$19.93 \\
   &         &         &$\pm$0.24  &$\pm$0.30&$\pm$0.08 &$\pm$0.12 & & &$\pm$0.10&  $\pm$0.19&  \\
J1(*) & 27.39   & $>$27.2 & $>$28.1   & $>$27.5 & 25.21    & 24.77& 24.82& 24.80 &21.96& 22.23 &$>$19.93 \\
    &$\pm$0.49&         &           &   &$\pm$0.08 &$\pm$0.11    &$\pm$0.25   &$\pm$0.22 &$\pm$0.18 &$\pm$0.21&  \\
\hline
\end{tabular}
\tablefoot{Information given in this table: (1) Object Identification, AUTO magnitudes in filters $I$ to $Ks$ (columns
  2 to 7), corrected 2\arcsec \ diameter aperture magnitudes for IRAC 3.6,
  4.5, 5.8 and 8.0$\mu$m (8, 9, 10, 11), and MIPS 24$\mu$m (12). 
(*) These objects are also non-detected on the 13.2ksec $z$-band FORS2 image
  centered on the cluster core.  
Non-detections are indicated as $1 \sigma$ upper limits.}
\end{table*}
\section{Selection of high-$z$ galaxy candidates}
\label{selection}

    The original catalog on the HAWK-I field of view includes $\sim
5\times10^4$ objets over a $\sim$45 {\rm arcmin}$^2$ area with more than 75\%
exposure time on the HAWK-I images. Since high-$z$
candidates should be detected both in the $H$ and $Ks$ bands, we first require a
detection above 5$\sigma$ level in both filters within a 1.3\arcsec \ diameter
aperture. This means selecting sources with $H<$ 26.22 and $Ks<$ 25.43,
corresponding to a completeness level in our survey of $\gtapprox$70\% for point
sources (see Fig.~\ref{completeness}). We also require less than a 2$\sigma$
level detection in both $I$ and $z$ bands. This selection yields a sample of
175 objects after removing all noisy areas on the $I$ and $z$ images, i.e. a
common clean area of $\sim$42{\rm arcmin}$^2$
(equivalent to $\sim$33{\rm arcmin}$^2$ after correction for lensing dilution). 

\begin{figure}[htb]
   \centering
\includegraphics[width=0.45\textwidth]{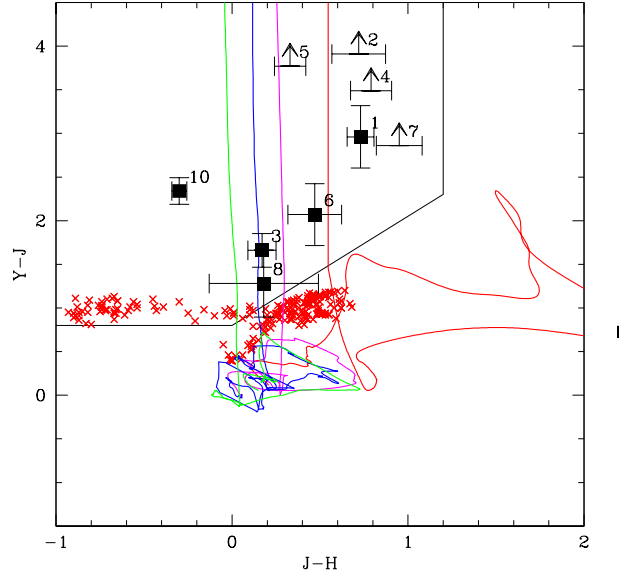}
\caption{Color-color diagram showing the position expected for spectral templates 
with redshifts $z \sim$ 0--9.0: E-type galaxies (CWW; red solid line),
Im-type galaxies (blue line), and starburst templates of Kinney et
al.\ (1996) (magenta and green lines). Red crosses show the colors of 
M, L, T, stars from the SpeX Prism libraries of Burgasser et al.\ (2006).
Black lines delimit our selection window for $z$ $\gtrsim$ 7.3 galaxies.
Our 9 $Y-$dropout candidates are marked with black squares and arrows:
\# 1 to \# 8 correspond to Y1 to Y8, whereas \# 10 is z1.}
\label{CC_YJH}
   \end{figure}

\begin{figure}[htb]
\centering
\includegraphics[width=0.45\textwidth]{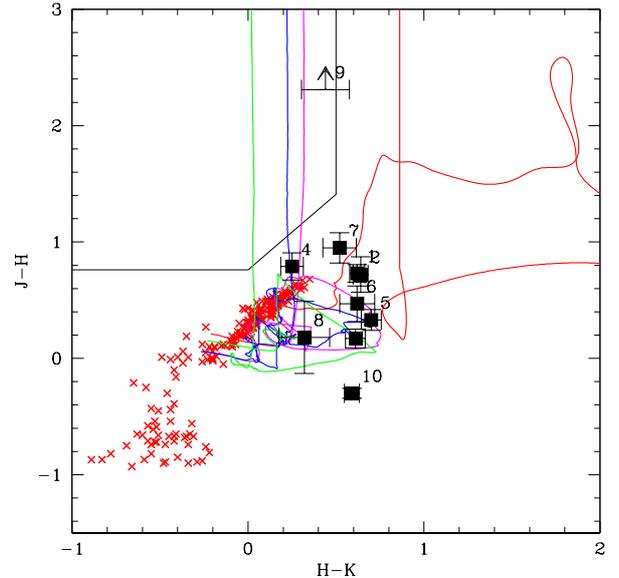}
\caption{Same as Fig.\ \protect\ref{CC_YJH} for (J-H) versus (H-K) colors
showing one $J-$drop candidate (J1, \#9) and the 9 $Y-$dropout candidates
(same identifications as in Fig.\ \protect\ref{CC_YJH}).
}
\label{CC_JHK}
\end{figure}

\begin{figure}[htb]
\centering
\includegraphics[width=0.45\textwidth]{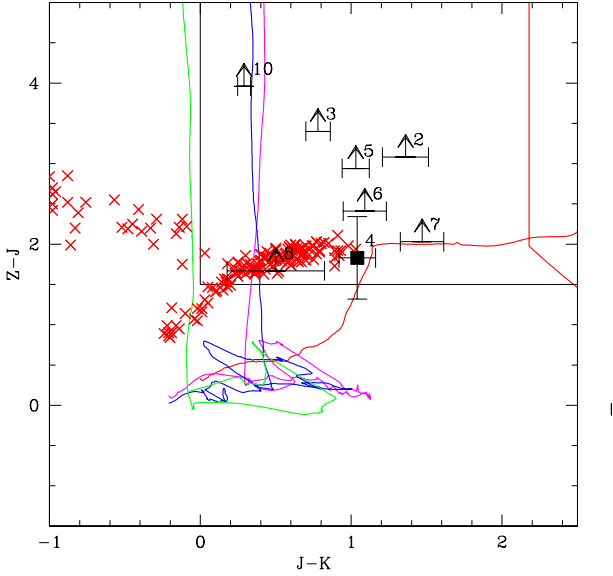}
\caption{Same as Fig.\ \protect\ref{CC_ZJK} for (z-J) versus (J-K) colors.
Thin black lines denote the selection window (c) (the same as in Capak et
al.\ 2011). 
The same identifications as in Fig.\ \protect\ref{CC_YJH} and
Fig.\ \protect\ref{CC_JHK} are used for the dropout candidates. 
}
\label{CC_ZJK}
\end{figure}

We then apply the following criteria to the remaining sample 
in order to select objects at $z>7$:

\begin{itemize}

\item [(a)] $Y-J>0.8$, $J-H<1.2$, and $Y-J>1.25\times(J-H)+0.8$. 
This window, based on simulations using different spectral templates, selects 
$Y$-dropout $z\ge7.3$ candidates. Fig.~\ref{CC_YJH} displays the
corresponding color-color diagram for different models, namely 
E-type galaxies (CWW), Im-type galaxies (constant star-formation model from
Bruzual \& Charlot 1993, 2003), and starburst templates of Kinney et al.\ (1996),
together with the selection window. 
Other galaxy templates, such as the UV-to-radio spectral templates of galaxies 
and AGN from Polletta et al.\ (2007) used subsequently not shown here, populate 
a very similar area is this and other color-color plots.
Also shown are the synthetic colors of cool stars (M, L, T types) from the 
SpexGrism spectral library (see Burgasser et al.\ 2006, and references therein)
\footnote{See {\tt http://www.browndwarfs.org/spexprism}}.

\item [(b)] $J-H>0.76$, $H-Ks<0.5$, and $J-H>1.3\times(H-Ks)+0.76$
illustrated in Fig.\ \ref{CC_JHK}.
Furthermore we require a non-detection in the $Y$ band, below the 2 $\sigma$
limit. 
This window is intended to select $J$-dropout candidates in the range $8\lesssim z
\lesssim 10$ wich are well detected in $H$ and $Ks$ bands (cf.\ Richard et al.\ 2006). 
The corresponding color-color plot is shown in Fig.~\ref{CC_JHK}.

\item [(c)] The $z$-dropout selection criteria adopted by Capak et al.\ (2011) for
their sample, i.e. $J<$ 23.7 (their 5$\sigma$ detection level), $z-J\ge1.5$,
$J-Ks>0$, and $Ks-4.5\mu m >0$ (cf.\ Fig.\ \ref{CC_ZJK}).
The last criterium involving the $4.5\mu m $
band was not applied to our sample given the partial coverage of the HAWK-I
field of view. Fig.~\ref{CC_ZJK} displays the corresponding color-color plot. 
This selection does in particular not make use of the $Y$ band filter,
intermediate between $z$ and $J$, available for our observations.

\end{itemize}

   There is some overlap between the $z$ and $Y$ bands, leading to somewhat
ill-defined dropout criteria in $z-Y$. For this reason, we use instead $Y-J$
and $z-J$ in the above selection windows. Including the $Y$ band provides us
with a useful discrimination between high-z galaxies and cool stars, as shown
in Fig.~\ref{CC_YJH}, while improving photometric redshifts. 

%

%

   The blind selection windows (a) and (b) include 48 and 52
sources respectively (25 sources in common), i.e.\ 75 candidates in total out of the initial
sample of 175 optical dropouts. Note that this
selection does not introduce any restriction in magnitude except for the
$\ge$5$\sigma$ level detection in $H$ and $Ks$. When applying the $z$-dropout
selection criteria (c), 8 sources are found. All these optical dropouts were carefully
examined by a visual inspection in the different (original) bands in order to
reject both spurious detections in 
the near-IR images and false non-detections in the $I$ and $z$ bands. The main
sources of contamination were images detected within the haloes of bright
galaxies leading to fake measurements or highly contaminated photometry. A mask
was created to remove these regions from the subsequent analysis (this
corresponds to $\sim$15\% of the total surface). 

We have also checked that the selection based on the detection in $H$ and
$Ks$ bands does not introduce a bias against intrinsically blue sources (see
e.g. Finkelstein et al.\ 2010). The selection was repeated with the
requirement of a detection above 5$\sigma$ in the $H$-band alone, all the
other conditions for optical dropouts being the same. This new selection
includes all the previous objects plus four additional sources, but none of
them survived the manual inspection.

   At the end of the visual inspection, only 10 candidates survive
from the original sample. Their photometry is listed in Table \ref{tab_catalog2}.
The brightest one in $J$ is the only source retained by selection criteria (c) 
after manual inspection (identified hereafter as z1). This source
and 8 additional objects fulfill the $YJH$ color selection (a) 
(cf.\ Fig.~\ref{CC_YJH}), which closely follows the selection previously
adopted by Bouwens et al.\ (2008) and Richard et al.\ (2008) in $zJH$.
The latter 8 objects, denoted as Y1 to Y8, are listed 
in order of increasing $H$ band magnitude within 1.3\arcsec aperture
(cf.\ Table \ref{tab_catalog2}).
Subsequently we refer to these 9 sources as $Y-$drop candidates. 

   Only one candidate, J1, fulfills the selection criteria (b) based on $JHKs$,
and is also consistent with the z$\sim$9 $J-$drop selection criteria by
Bouwens et al.\ (2010) (see Fig.~\ref{CC_JHK}). This object was 
formally detected in the $I$ band at \ltapprox2$\sigma$ level (double-image
mode), as indicated in Table \ref{tab_catalog2}, although no
convincing counterpart is seen in this image (see also Fig.~\ref{trombino}).
This source has a counterpart detected on the HST F850LP/ACS image, 
as discussed in Sect.~\ref{observed} below. 

   It is worth noting that all $Y$-dropout candidates excepted z1 are too faint
to be selected by the original Capak et al. criteria (i.e. intrinsic
lensing-corrected $J$\gtapprox 24, see below). 
When the same color selection (c) is applied to fainter $J$-band magnitudes, up to
$J<$ 25.7 (our own 5$\sigma$ detection level), 22 objects are selected after
manual inspection. Among them, all our Y candidates except Y4 are found. 
Indeed, Y4 was formally detected in the reference $z$ band at
\ltapprox2$\sigma$ level (as indicated in Table \ref{tab_catalog2},
double-image mode), although there is no clear counterpart seen on this image and
it is also not detected in the $z_2$ field. 
This sample of fainter $z$-dropout candidates will be discussed elsewhere in a
forecoming paper.  

   Hereafter we concentrate on the bright $z$-dropout z1 and the nine $Y$ and
$J$-dropout candidates. We have checked for all these objects that their
spectral energy distribution (hereafter SED) remains the same when using
SExtractor aperture and MAG\_AUTO magnitudes. MAG\_AUTO magnitudes were
preferred for the subsequent analysis because they are closer to total
magnitudes.   

\section{Results}
\label{results}


\subsection{Observed properties of the high-$z$ candidates}
\label{observed}

   The identification, position and morphology of the ten bright $Y$ and
$J-$dropouts selected in this field is given in Table~\ref{tab_catalog1}.
Table~\ref{tab_catalog2} summarizes their photometric properties, and
Fig.~\ref{trombino} displays the corresponding postage stamps.  
Except for J1, all candidates are detected in the $J$, $H$ and $Ks$ bands. 
The observed $H$ and $Ks$ band magnitudes of our objects range typically from 
$\sim$ 23 to 25, i.e.\ in a regime where our sample is close to 100\% complete. 
For the high-$z$ candidates in the common area between $z_1$ and $z_2$ fields
of view (namely z1, Y3, Y4, Y5, Y6, Y8 and J1), we have also cheched the
independent non-detection on the two original images. 

   The SExtractor stellarity index 
\footnote{This index ranges between 0.0 for extended sources and 1.0 for
unresolved ones (see Bertin \& Arnouts 1996).}
and the FWHM along the major axis, both
measured on the $H+Ks$ detection image, were used to quantify the morphology
of our candidates. This information was only used to assess the possible
contamination by cool stars, in addition to colors. The reliability of the 
SExtractor stellarity index for galaxies diminishes towards the faintest magnitudes. 
We have checked that a reliable index can be obtained up to $HK \sim$24.5 for
our combined image, becoming hazardous for fainter sources and highly
unreliable at $HK\ge$25.0, where the S/N significantly drops below
$\sim$10. From Table~\ref{tab_catalog1} it appears that all our sources seem
to be inconsistent with stars, except Y8 and J1 which are too faint for a
robust morphological classification based on the detection image.  
As shown in Fig.~\ref{CC_YJH} to ~\ref{CC_ZJK}, all candidates but Y8
display colors which are clearly incompatible with cool stars. 

   J1 is the only candidate located on the central area covered by the HST
F850LP/ACS image. A faint and compact object is indeed detected on this image,
with FWHM$\sim$0.1\arcsec (Sextractor stellarity parameter is 0.7), 
and $z_{850}=$27.39$\pm$0.18 (MAG\_AUTO on the
original HST image). This magnitude is fully consistent with the non-detection
on the ground-based $z$-band images (AB$\sim$27.2 at 1$\sigma$ level), and
confirms the important break between optical and near-IR bands for this
object, which is the faintest one in our sample. 
 
   Good quality magnitudes were extracted for six and seven sources on the IRAC
3.6$\mu$m and 4.5$\mu$m images respectively. For the remainder, only partial data is
available (Y1, Y2), or the photometry is strongly contaminated by nearby sources (see
Fig.~\ref{trombino}). The only candidates clearly detected and reported in
Table~\ref{tab_catalog2} for the shallow 5.8$\mu$m and 8.0$\mu$m images are 
z1, Y8 and J1, even though magnitudes are dubious for Y8 and J1 due to a noisy
environment. 
Two objects, z1 and Y7, are clearly detected on the MIPS 24$\mu$m image. 


\subsection{Magnification of the drop-out sources}
\label{magnification}

    The lensing model for A2667 was originally obtained by Covone et
al.\ (2006). We use this model to compute the magnification maps at different
source redshifts with the public lensing software Lenstool\footnote{{\tt
http://www.oamp.fr/cosmology/lenstool}}, including the new MCMC
optimization method (Jullo et al.\ 2007) providing bayesian estimates on model 
parameters.  

    The mass model was used to derive the
magnification factor for each object and associated error bars, both for the
high and low-z solutions (i.e. typical redshifts of $z\sim7.5$-9.0 and
$z\sim1.7$-2.0 respectively, see Sect.~\ref{photoz}). 
These values are given in Table~\ref{tab_catalog1}. 
The magnification factors of our objects typically range between 1.1 and 1.4, 
e.g.\ $\sim$0.4 mags being largest for Y4 and J1. None of
these candidates is expected to be a multiple image. 

Given the location of our candidates with respect to the critical lines,
either in the high or in the low-z solutions,
the uncertainty in the magnification associated to the uncertainty on
the redshift is smaller than 10\% in all cases. Also the magnification factor 
at a given position on the image plane varies slowly with redshift for sources
located more than $\sim 10''$ apart from the critical lines, and this is
indeed the case for all our candidates. 
Error bars in magnification are also given in Table~\ref{tab_catalog1}, 
including both uncertainties on source redshift and systematic errors due to
the choice of the parametrization in lensing modelling (see e.g. Maizy et
al.\ 2010).  
We also used this lensing model to compute the effective surveyed area and
volume around $z\gtrsim$7.5 when comparing with blank field surveys. All
surface and volume number-densities given in this paper have been corrected
for magnification by the lensing cluster.

\begin{figure*}
\centering
\includegraphics[angle=270,width=0.80\textwidth]{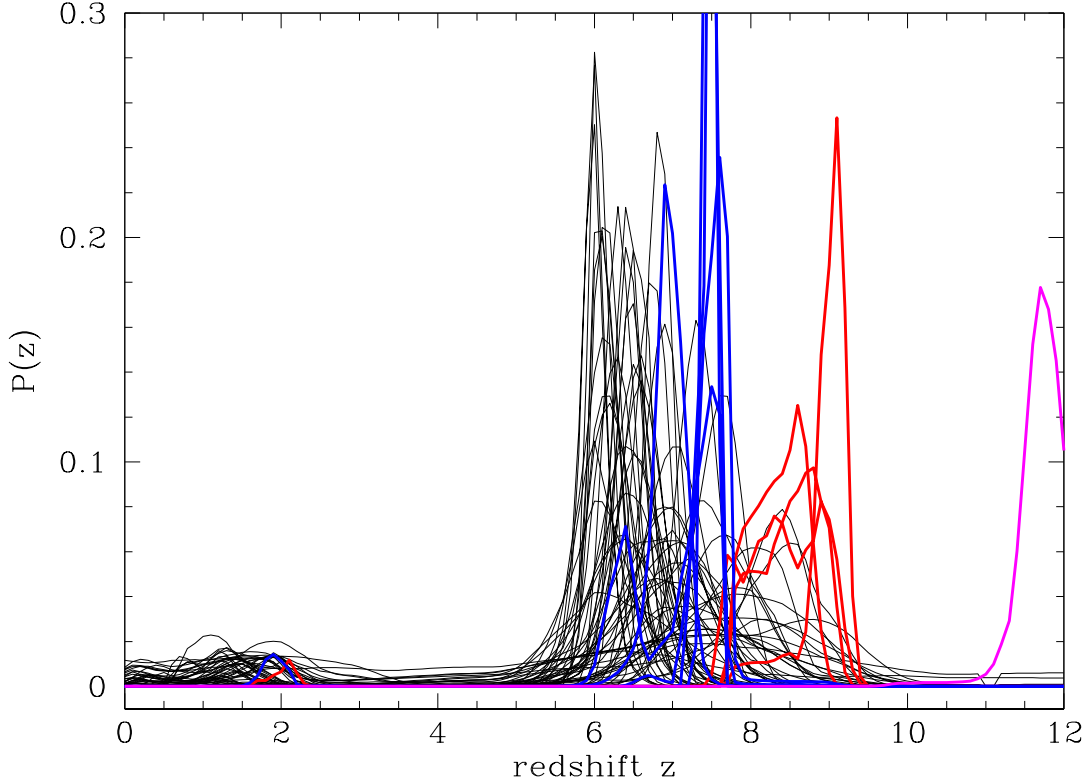}
\caption{
Redshift probability distributions, $P(z)$ for our objects computed using standard templates 
(BC solar metallicity, no emission lines).
Objects Y1, Y3, Y6, Y8, and z1 are shown in blue, Y2, Y4, Y5, and Y7 in red, and
J1 in magenta.
Black lines show $P(z)$ of other $z \sim$ 6--8 galaxies
from HST surveys using NICMOS and WFC3 from the analysis of
Schaerer \& de Barros (2010), namely Gonzalez et al.\ (2010) and
McLure et al.\ (2010) sample)}
\label{fig_pz}
\end{figure*}
%


\subsection{SED fitting results: photometric redshifts}
\label{photoz}

Photometric redshifts and associated probability distributions were
derived for each one of our candidates from the available broad-band
photometry, except for the 24\micron\ MIPS photometry, which is only available
for a few candidates (see discussion below in Sect.~\ref{properties}). 
A modified version of the public photometric redshift software $Hyperz$
(Bolzonella et al.\ 2000) was used, adapted to
include nebular emission (see Schaerer \& de Barros\ 2009, 2010,
hereafter SB2010).  
The following spectral templates were used for our SED fits: empirical
templates (starbursts from Kinney et al.\ 1996; galaxies from Coleman
et al.\ 1980; GRASIL templates from Silva et al.\ 1998,
and the UV-to-radio templates of galaxies
and AGN from Polletta et al.\ 2007), and Bruzual \& Charlot (2003; hereafter
BC) evolutionary sythesis models to which nebular emission (lines and
continua) is added optionally.

The free parameters for the SED fits are the metallicity $Z$ (ranging
from $\sim $1/50 $Z_{\odot}$ to $Z_{\odot}$), the star-formation
history, the age since the onset of star-formation, extinction and
redshift. For empirical templates redshift is in principle the only
free parameter. However, in some cases we also allow for additional,
variable extinction for empirical templates.  Extintion is varied from
$A_V=0$ to 4 in steps of 0.2 mag, using the Calzetti et al.\ (2000)
extinction law. The Lyman forest blanketing is included following the
prescription of Madau (1995). 

The non-detection in the $I$, $z$, $Y$ and $J$
bands was used as a constraint when computing photometric redshifts. 
Unless otherwise indicated, the rule applied corresponds to the usual 
case ``1'' of $Hyperz$, i.e. the flux in these filters is set to zero, 
with an error bar corresponding to the flux at 1$\sigma$ level, using 
both the global value and the local sensitivity computed
near each source (see below). We have also computed photometric redshifts
by forcing the fluxes in these filters to be below 2$\sigma$ and 3$\sigma$
levels using case ``2'' of $Hyperz$ for non-detections.
As explained below, the results obtained are
rather insensitive to the non-detection rule and sensitivity applied (local or
global).  
A minimum error bar of 0.1 magnitudes was assumed for IRAC photometry to
account for uncertainties in absolute flux calibrations when combining with
the other filters. 

The resulting redshift probability distributions $P(z)$ for all our objects,
obtained with standard Bruzual \& Charlot, solar metallicity
templates, is shown in Fig.~\ref{fig_pz}.  
$P(z)$ displayed in this Figure were obtained using the $Hyperz$ approach,
i.e. $P(z)\propto exp(-\chi^2(z))$,
which is very similar to the results derived from Monte Carlo simulations.
As shown by this Figure,
most objects have a relatively well defined redshift probability
distribution $P(z)$ peaking at high redshift. Five objects (Y1, Y3,
Y6, Y8, and z1) show best-fit redshifts \zphot$ \sim$ 7--8, four
objects (Y2, Y4, Y5, and Y7), a higher redshift \zphot$ \sim$
7.5--9.5, and for the $J-$drop J1 \zphot$>$9.5 is favoured. These
redshift ranges and their relative grouping are consistent with
expectations from their colors (cf.\ Fig.~\ref{CC_YJH}).  
For most objects a less significant solution is found at low-$z$, 
in general within $z\sim$ 1.7 and 2.8.
However, even though the high-$z$ solution produces a better fit, 
several of these sources seem too bright (M$_{1500}<-23.0$)
to be at $z\ge7.5$, suggesting some contamination by low-$z$ interlopers. This
issue will be discussed below. 

    Fig.~\ref{fig_pz} also displays for comparison the $P(z)$ derived 
by SB2010 
for a recent sample of $z\sim$ 6--8 galaxies
including objects from HST surveys using NICMOS and the recent WFC3 camera. 
The corresponding $P(z)$ of our candidates compare to and overlap these
samples, and our selection function has clearly favoured the $z>$7 domain. 

\begin{figure*}
\centering{
\includegraphics[width=0.33\textwidth]{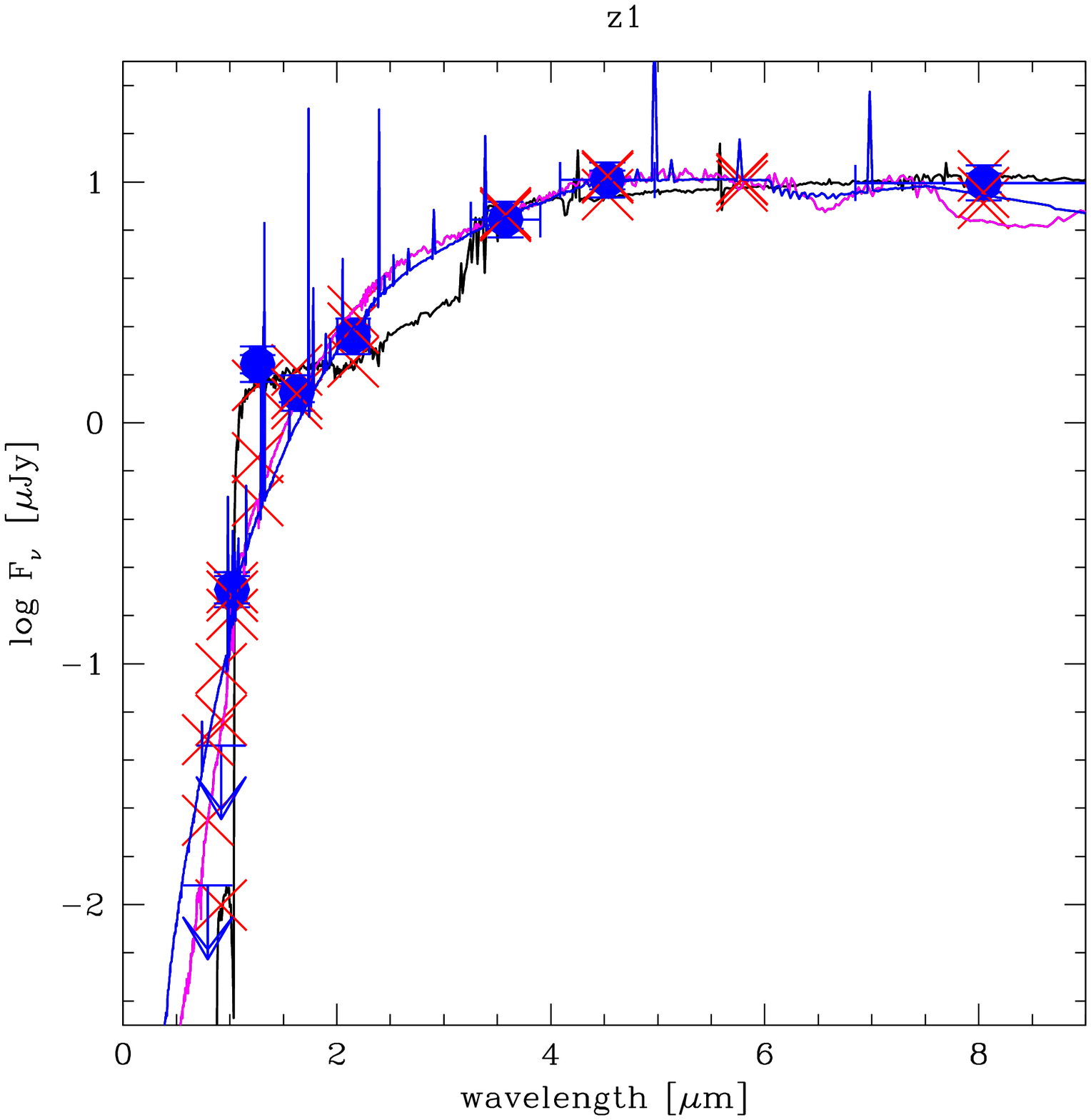}
\includegraphics[width=0.33\textwidth]{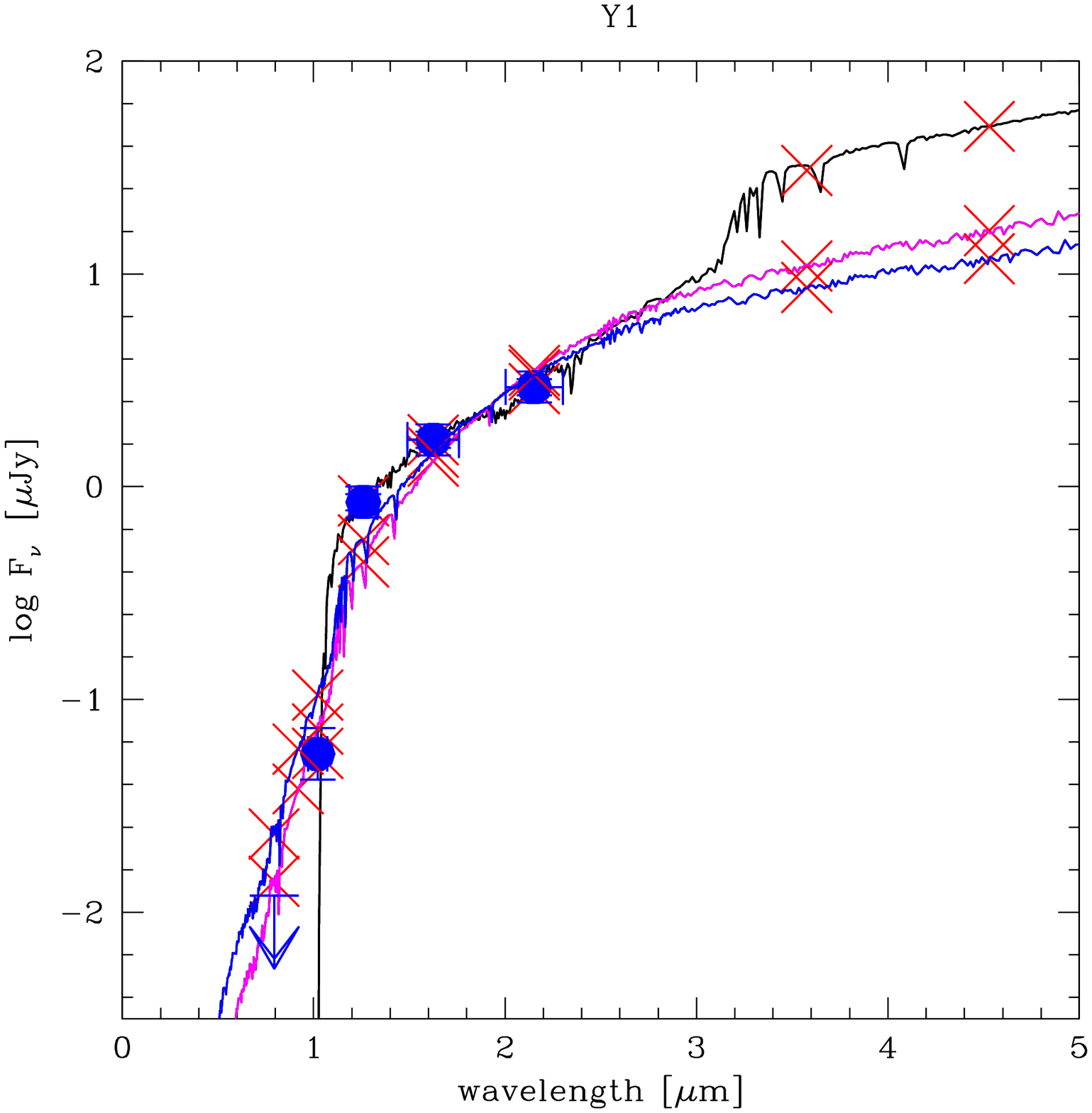}
\includegraphics[width=0.33\textwidth]{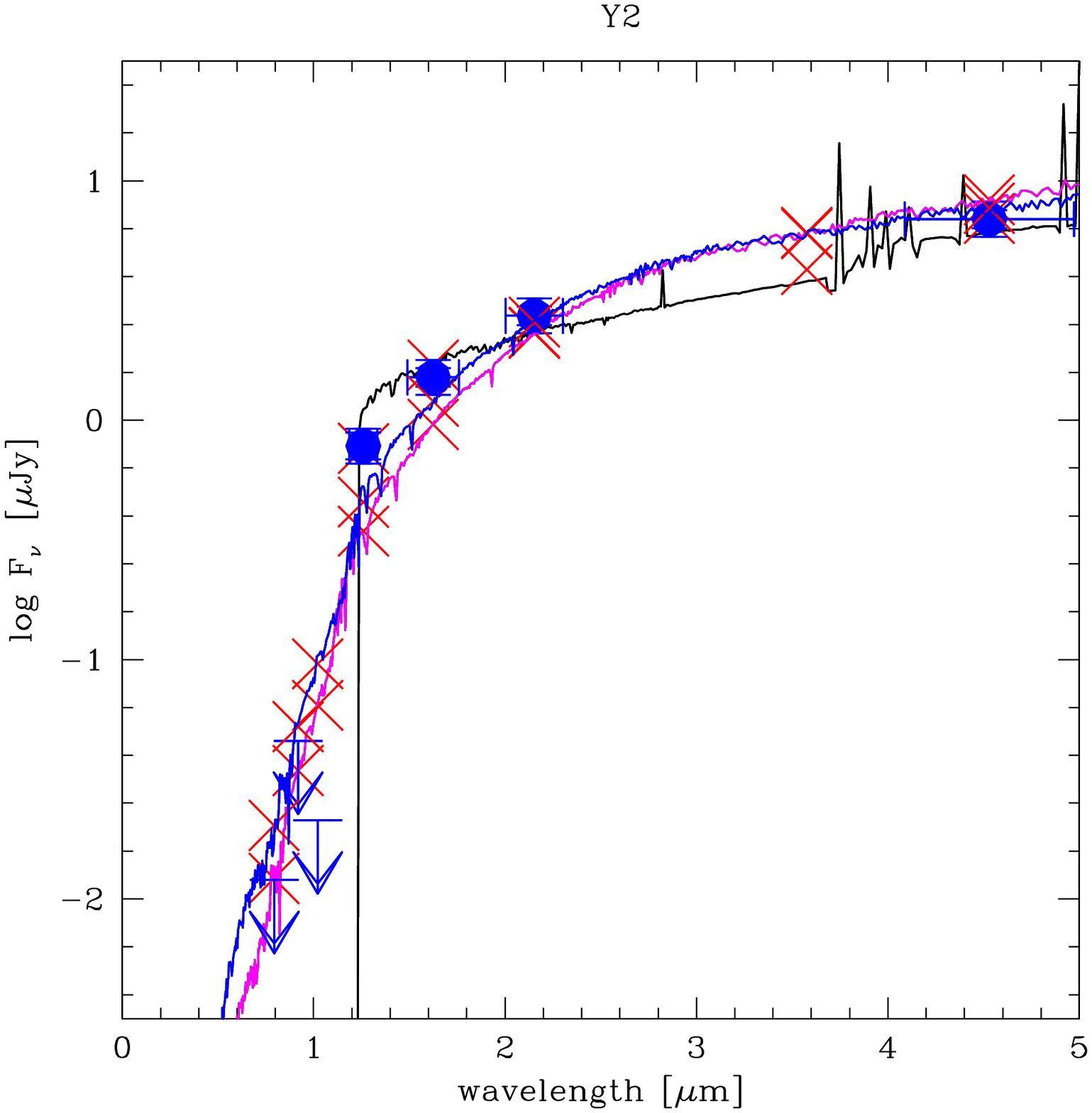}
}
\centering{
\includegraphics[width=0.33\textwidth]{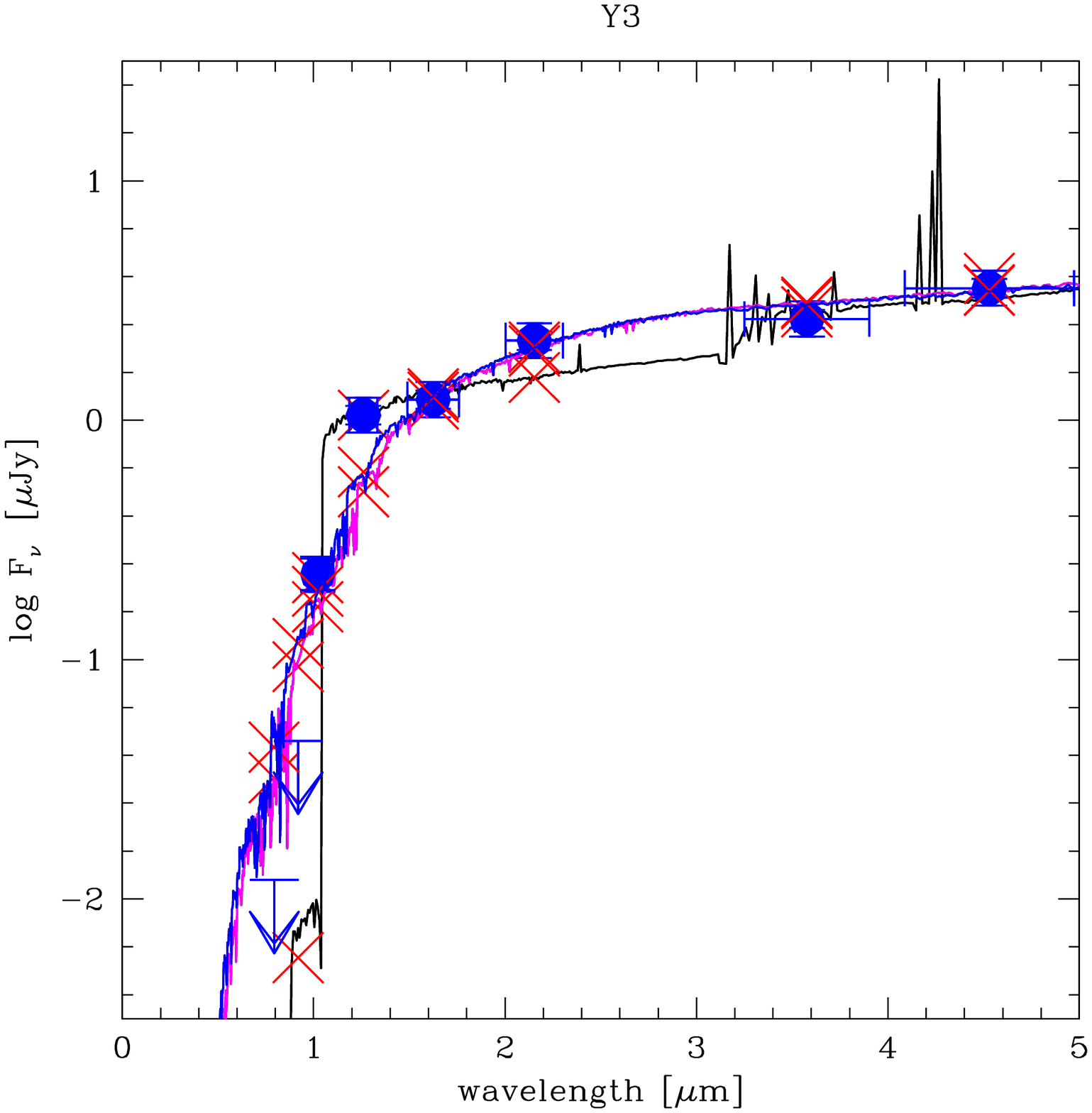}
\includegraphics[width=0.33\textwidth]{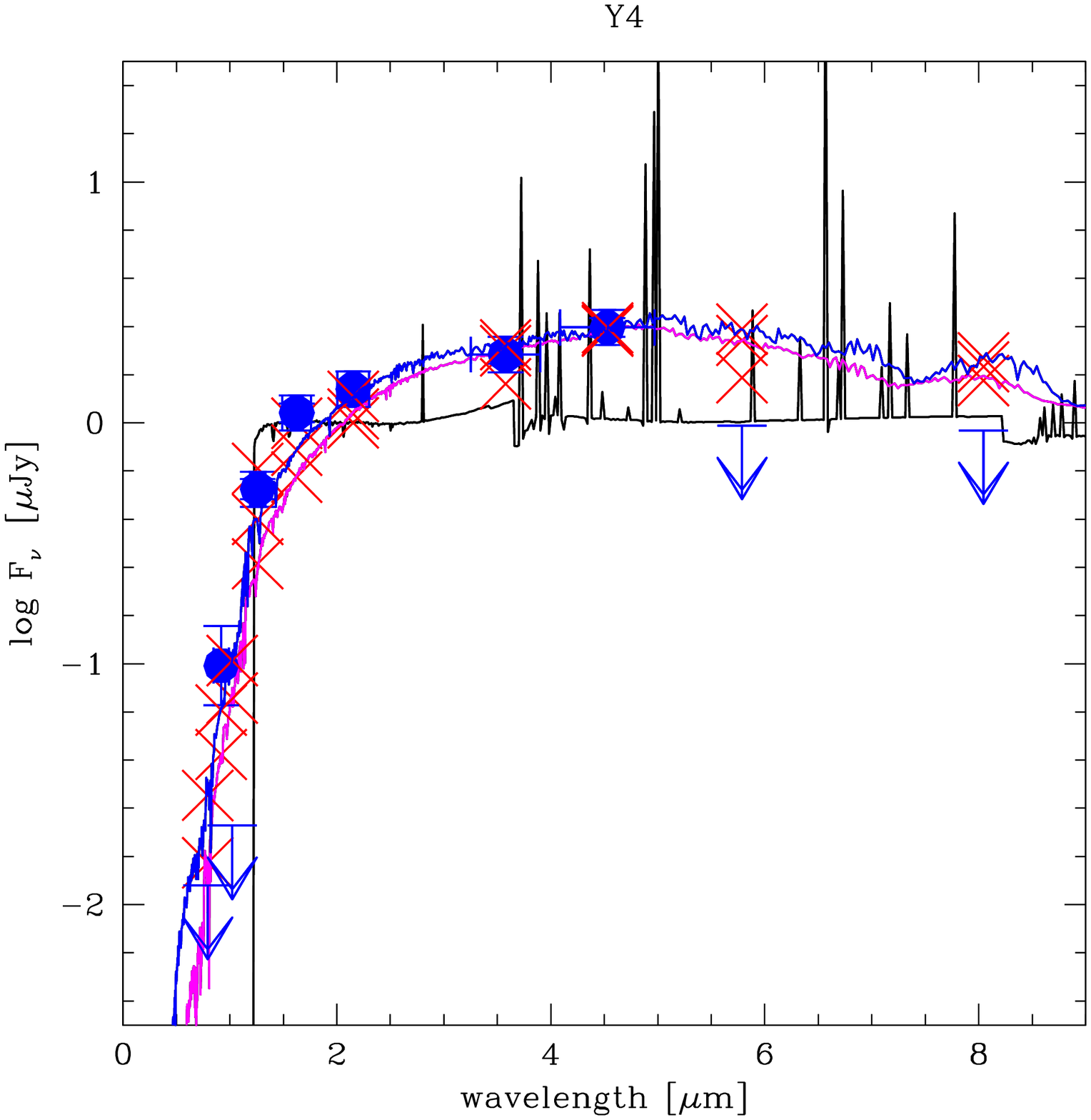}
\includegraphics[width=0.33\textwidth]{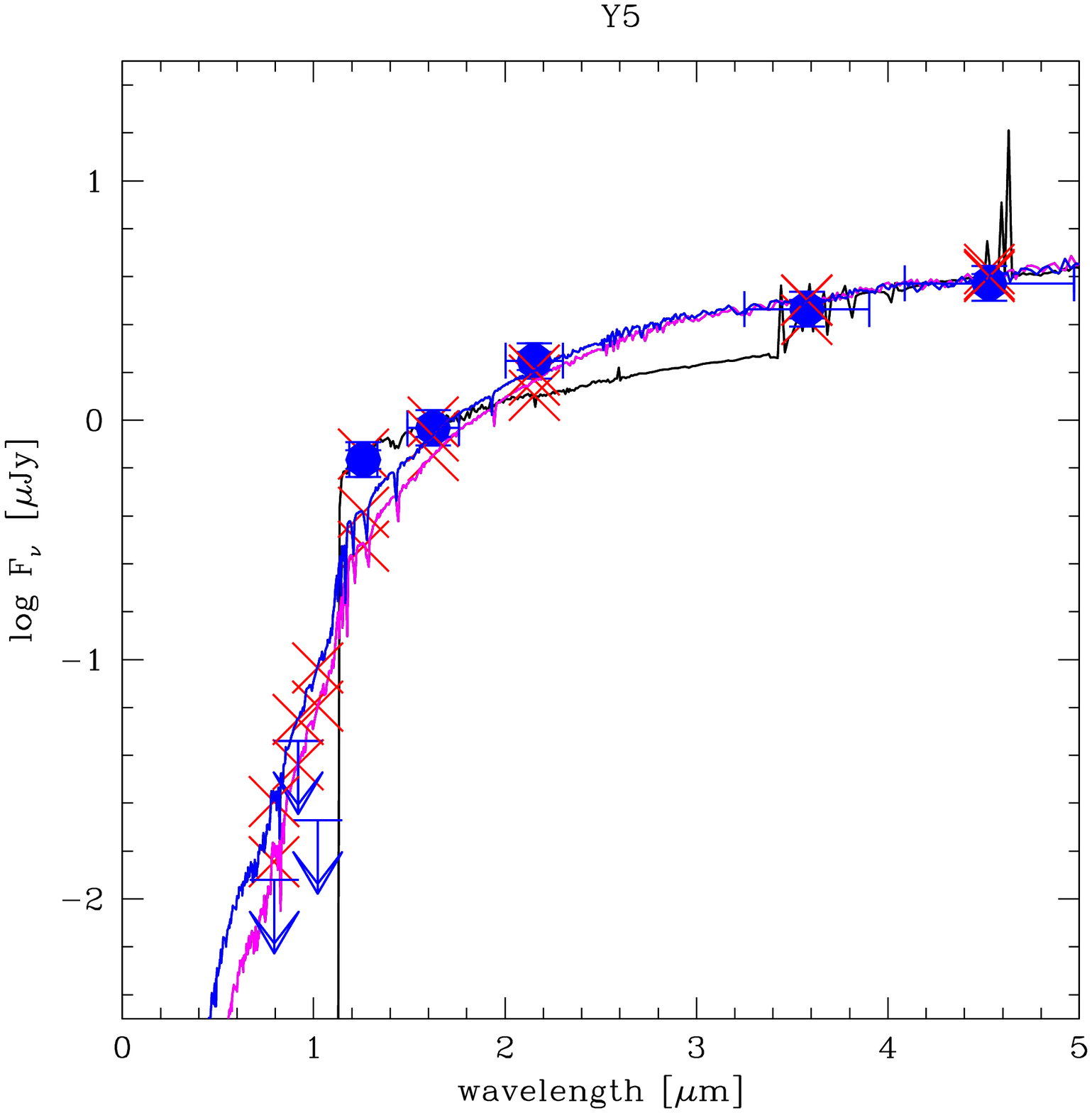}
}
\centering{
\includegraphics[width=0.33\textwidth]{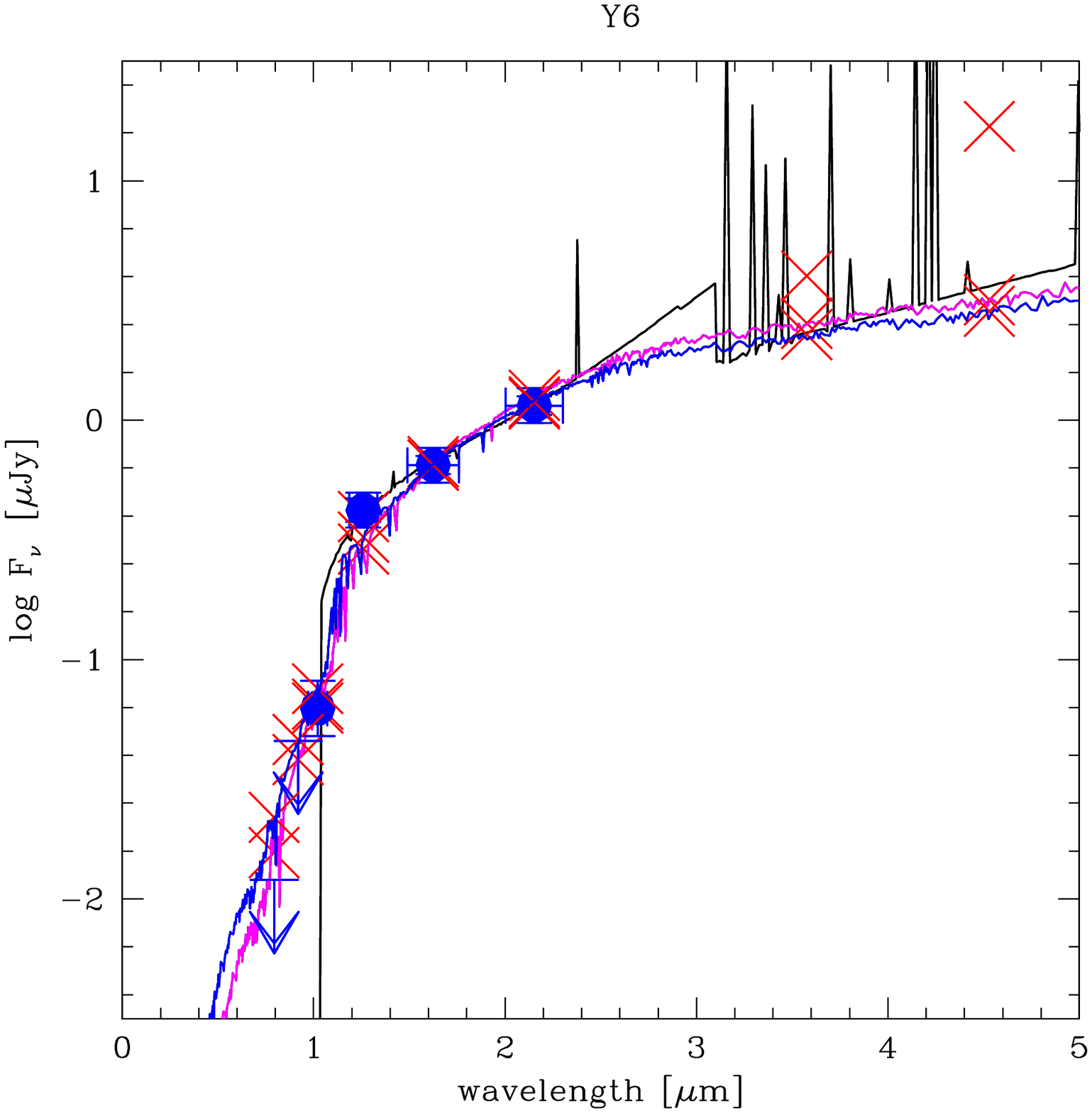}
\includegraphics[width=0.33\textwidth]{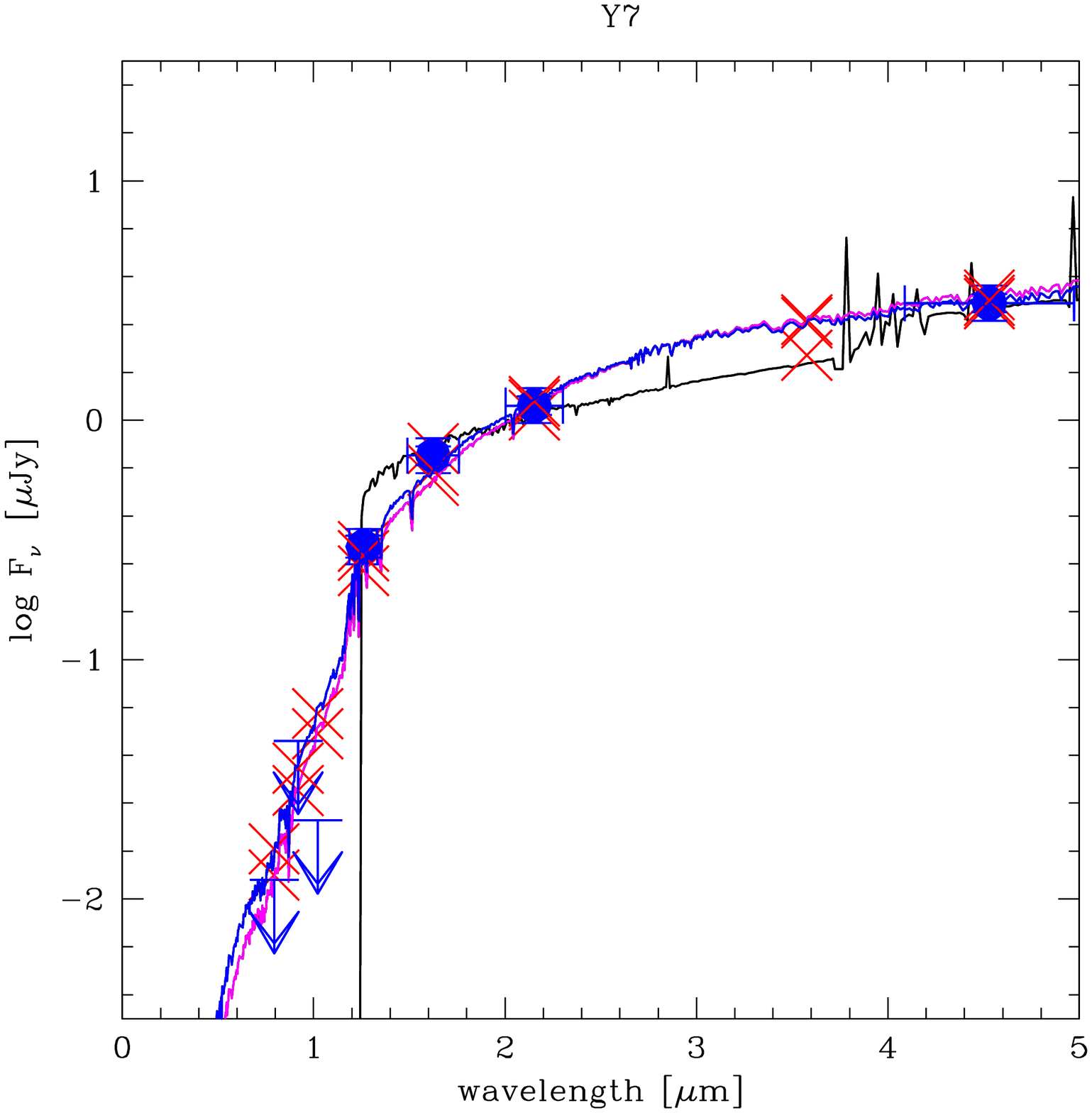}
\includegraphics[width=0.33\textwidth]{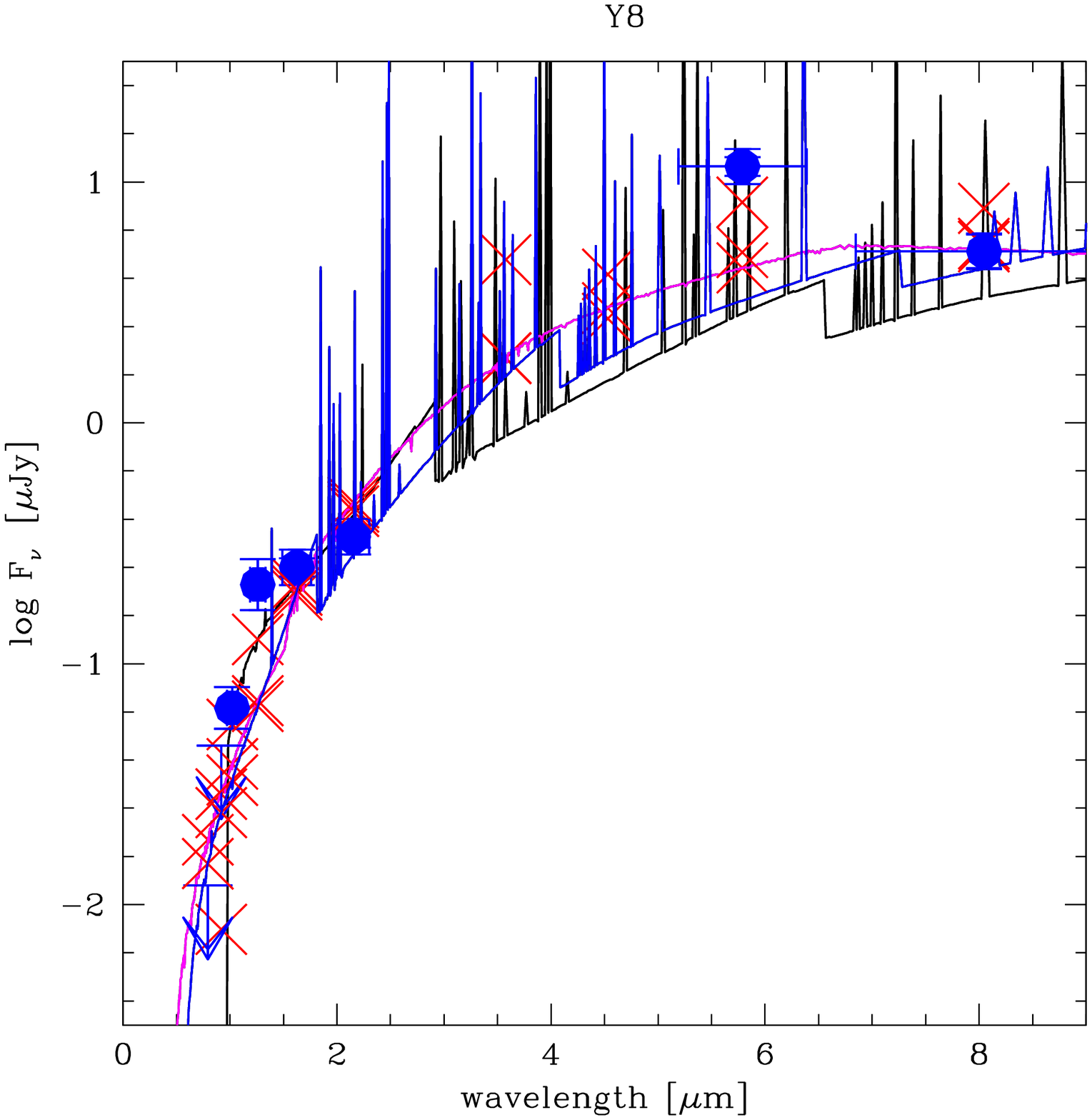}
}
\centering{
\includegraphics[width=0.33\textwidth]{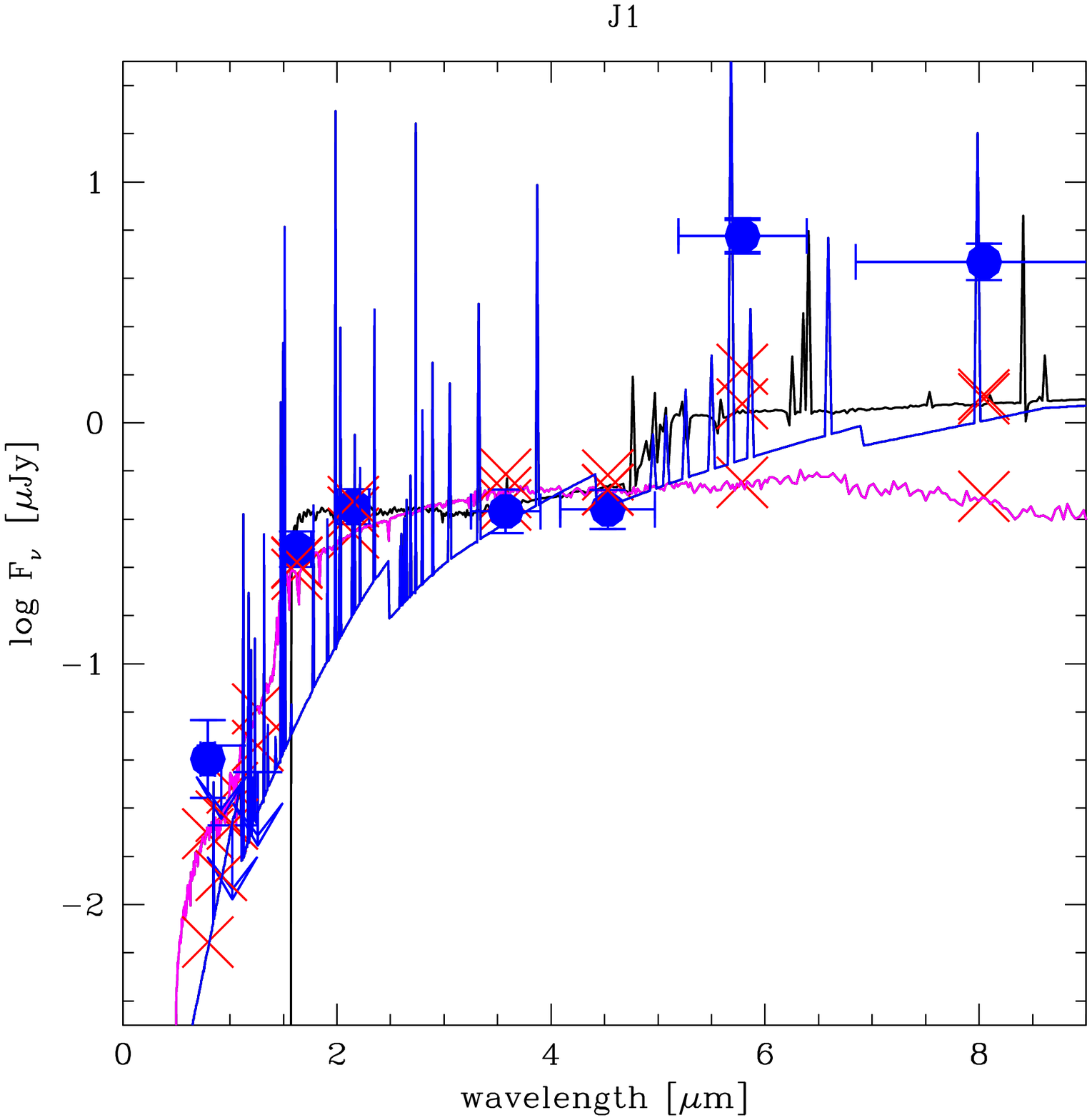}
}
 \caption{Best-fit SED solutions for high-$z$ (black lines) et low-$z$ (blue and
magenta lines) for all the bright $z$, $Y$ and $J-$dropouts found in A2667. 
Error bars and upper limits correspond 
to 1$\sigma$ values, as reported in Table~\ref{tab_catalog2}. 
Red crosses indicate the synthetic flux in the filters.
High-$z$ (black) spectra include nebular emission. Low-$z$ solutions are
displayed for the complete library (including nebular emission) in blue, and
for the standard templates in magenta. 
Note the extended wavelength scale for objects with available 5.8 and 8.0
\micron\ photometry.
}
\label{SED-fit}
\end{figure*}

Using standard spectral templates (neglecting the effects of nebular 
emission) we obtain the best-fit photometric redshifts and physical 
parameters given in Table~\ref{tab_prop}.
We have also examined how the inclusion of nebular lines and 
continuous emission may alter the photometric redshift.
In Fig.\ \ref{SED-fit} and Table~\ref{tab_prop}
we show the best-fit SEDs for all our objects in two
different cases: without any redshift prior, and with the restriction of $z<4$. 
In all cases the best-fit is found at high $z$, irrespective of the inclusion
or not of nebular emission. Low redshift solutions (typically at $z \sim$ 1.7--2.1
and $z \sim$ 2.5--2.8 for J1),
show fits of lower quality, especially close to the spectral break
and between $Y$, $J$, and $H$, as could be expected from the behaviour
of spectral templates in these colours.
Furthermore, these best-fit, low redshift solutions show generally 
excess in the optical bands ($I$, $z$) indicating that most of these objects
should be detected at a 3--5 $\sigma$ level in at least one of these bands.
Note, that in some cases the inclusion of nebular emission allows for 
somewhat "unexpected" solutions with strong emission lines plus 
a high attenuation (see e.g.\ the fits for z1, Y8, and J1), although the 
resulting low-$z$ fits remain with a higher $\chi^2$.

We have investigated the influence of non-detection rules and limiting fluxes on 
photometric redshifts results. When applying a local non-detection limit
instead of the global one, there is no difference in the high-$z$ solutions
(similar $\chi^2$ and $\Delta$z$<$0.1). Low redshift fits display the
same lower quality as compared to high-$z$, with similar $\chi^2$ and, in
general, $\Delta$z$\sim$0.1, although there are larger differences for
Y2 ($\Delta$z$\sim$1), Y4 ($\Delta$z$\sim$0.4) and Y5 ($\Delta$z$\sim$0.2).
As shown in Table~\ref{tab_prop},
the same differences in $\Delta$z are observed when comparing the photometric redshifts 
achieved with local non-detections and the standard $Hyperz$ models, 
with those found with global non-detections and the complete library
(including nebular emission). In other words, the dispersion in photometric
redshifts between local and global non-detection limits is similar to the
dispersion due to model uncertainties. And, in all cases, the high-$z$
solution is privileged. 

   We have also computed photometric redshifts by replacing 
the non-detection rule ``1'' of $Hyperz$ by rule ``2'' in all filters where
the candidates are formally not-detected,
where the flux and the error bars are set to $F_{\rm lim}/2$, for two
different cases, $F_{\rm lim}$=2$\sigma$ and 3$\sigma$ detection
levels. Best-fit redshifts remain precisely the same for most of our
candidates at 2$\sigma$ level, the only exception being Y8 (degenerate
solution with best fit at z$=$1.7). When the fluxes are allowed to reach a
3$\sigma$ level, three other objects become degenerate, with a best-fit at
low-$z$, namely Y6, Y7 and J1. 
Table~\ref{tab_prop} summarizes the integrated probability distribution at
z$>$6 for all candidates when using different assumptions for non-detections. 

   As mentioned above, there is some overlap between the $z$ and $Y$ bands,
but the $Y-$band allows us improving photometric redshifts. Indeed, when 
SED fitting results are derived without $Y-$band data, we still obtain a
best-fit at high-$z$ for all objects. The $P(z)$ distribution for all objects
detected in $Y$ (blue lines in Fig.\ \ref{fig_pz}) becomes broader (from
\zphot\ well peaked at 7--8 to \zphot $\sim$ 6.5--9). For all other objects
the changes in $P(z)$ are minor. 
When both $Y$ {\em and} $J-$band data are removed, this strongly degrades the
photometric redshift $P(z)$ distribution and \zphot\ becomes basically
undefined for all objects. 

\begin{table*}
\caption{\label{tab_prop}Properties of bright $z$, $Y$ and $J-$dropouts in A2667.}
\begin{tabular}{r|ccccccc|cccc|p{0.35cm}p{0.35cm}p{0.35cm}p{0.35cm}p{0.35cm}}
\hline
Source & $z_{phot}$ & $\chi^2$& $z_{min} - z_{max}$ & $A_V$ & M$_{1500}$ &
L$_{1500}$ & SFR &
$z_{phot}$ & $\chi^2$& $A_V$ & M$_B$ & P(z$>$6) &  &  &  &  \\
       &  high-$z$ &       &                &    &   &
$\times$10$^{41}$ &  &
low-$z$      &      &      &   & (1) & (2) & (3) & (4) &  \\
       &  &       &                &    &   &
erg/s/cm$^2$ & M$_{\odot}$/yr &
       &      &      &   &  &  &   &  &  \\
\hline
z1 (a) & 7.6 & 75.08& 7.5-7.7& 0.3 & -23.44 &13.7 & 144 &1.78 & 228.27 & 2.4 & -20.06 &1.0&1.0&1.0&1.0& (low)\\ 
(b)   & 7.6 &       &        &      &        &     &     &1.94 &      &     &    &  &  &  &  &  \\
Y1 (a) & 7.7 & 0.02 & 7.2-7.8& 2.4 & -23.12 &10.2& 108 &1.72 & 39.62 & 0.3 & -19.82 &1.0&0.0&1.0&1.0& low\\ 
(b)   & 7.4 &       &        &      &        &    &     &1.65 &      &    &  &  &  &  &  \\
Y2 (a)  & 8.7 & 0.89 & 7.7-9.3& 2.4 & -23.53 &14.9  & 157 &2.72 & 38.01 & 0.0 & -21.78 &1.0&0.0&1.0&0.99& low\\
(b)   & 9.1 &       &        &       &         &      &     &2.11 &      &     &    &  &  &  &   &   \\
Y3 (a) & 7.5 & 28.68 & 7.3-7.6& 1.2  & -22.97 & 8.9& 94&1.88 & 72.94 & 0.6 & -20.32& 1.0&0.97&1.0&0.79& high\\
(b)   & 7.5 &       &        &      &        &    &   &1.95 &      &     &    &  &  &  &    &  \\
Y4 (a) & 9.1 & 6.98 & 8.7-9.4& 1.2 & -23.14 & 10.4& 110& 2.58 & 73.71 & 0.0 & -21.13 &1.0&1.0&1.0&0.99& high\\
(b)   & 9.2 &       &        &     &        &     &    &2.11 &      &     &    &  &  & &   &    \\
Y5 (a)  & 8.6 & 10.82 & 7.7-8.8& 2.1 & -22.97 & 8.9& 94& 1.70 & 40.95 & 0.0 & -19.32&1.0&0.63&1.0&0.98& ?\\
(b)   & 8.3 &       &        &     &         &    &   & 1.94 &      &     &    &  &  &  &  &    \\
Y6 (a)  & 7.5 & 0.14 & 6.6-7.7& 1.2 & -22.15 & 4.2 & 44 &1.94 & 3.64 & 1.50 & -19.74 &0.94&0.0&0.55&0.46& low\\
(b)   & 7.5 &       &        &     &         &     &    &1.87 &      &     &   &  &  &  &   &   \\
Y7 (a)  & 9.1 & 0.05 & 8.0-9.4& 1.8 & -22.90 &8.3 & 87 &1.72 & 10.5 & 0.3 & -18.87&0.99&0.0&0.93&0.41& low\\
(b)   & 9.2 &       &        &     &         &    &    &2.11 &      &     &   &  &  &   &   &   \\
Y8 (a) & 7.4 & 0.02 & 5.9-7.7& 0.3  & -21.29 &1.9 & 20 &1.66 & 0.77 & 0.6 & -18.33&0.82&0.05&0.37&0.26& low\\
(b)   & 7.4 &       &        &     &        &    &    &1.70 &      &     &     &  &  &  &  &    \\
J1 (a)& 11.9 & 4.93 & 9.6-12.0& 0.0 & -22.66 & 6.7& 71 & 2.80 & 12.48 & 0.0 &-20.19 &1.0&0.94&0.85&0.20& (low)\\
(b)(c)  &  11.8 &       &        &     &       &    &    & 2.50 &      &     &  &  &  &  &   &   \\
\hline
\end{tabular}
\tablefoot{Information given in this table: (Column 1) object identification, \\
(2,3,4,5,6,7,8) best-fit photometric redshift at high-$z$, $\chi^2$, 1$\sigma$
confidence interval, best fit $A_V$, magnification corrected M$_{1500}$, 
L$_{1500}$ and SFR from Kennicutt (1998) calibration, \\
(9,10,11,12) best-fit photometric redshift at low-$z$ and corresponding
$\chi^2$, best fit $A_V$ and M$_B$, \\
(13,14,15,16) Integrated probability distribution for z$>$6,
normalized to 1, for different cases based on Bruzual \& Charlot models: 
(1) non-detection rule ``1'' of $Hyperz$, 
(2) the same for a P(z) including a luminosity prior
with non-detection rule ``2'' of $Hyperz$ (where the flux and the error
bars are set to $F_{\rm lim}/2$), and  
with $F_{\rm lim}$=2$\sigma$ (3) or
3$\sigma$ (4) detection level in all filters where the candidates are formally
not-detected.   \\
(17) Tentative classification 
between low and high-$z$ using a luminosity prior. In the case of
z1 and J1 (in brackets), the low-$z$ indentification is forced based on the
detections in the 24$\mu$m and $z_{850}$ bands respectively for z1 and J1
(see Sect.~\ref{properties} and Table~\ref{tab_catalog1}). \\
(a) Standard $Hyperz$ models (local non-detection limits). \\
(b) Complete library including nebular emission (global non-detection limits).  \\
(c) Photometric redshift for this source includes the detection in the
$z_{850}-$band. 
} 
\end{table*}


\subsection{Quality grades}
\label{grade}
 
   Table~\ref{tab_catalog1} includes 
two different quality grades for each source representing
its likelihood to be a genuine high-$z$ candidate. The first one is based on
the quality of the photometric information gathered for the source in terms of
surrounding environment, completeness of the SED and intrinsic UV luminosity
if at high-$z$. The second grade is based on the robustness of the optical
non-detection criteria, following Bouwens et al.\ (2010). 

   The first grade $Q$ includes three independent criteria
introduced as follows:

\begin{itemize}

\item  The first one ($Q1$) is the quality of the surrounding environment,
representing the possible contamination by neighbouring or underlying sources. Although
all candidates are isolated on the good-seeing detection image $H+Ks$, the
presence of another source within a distance of 2\arcsec is given the lowest
grade (=1), whereas isolated candidates without neighbours closer than
$\ge$3\arcsec have the highest grade (=3).  

\item The second one ($Q2$) is the quality of the photometric SED. Objects with
robusts constraints available beyond the $Ks$ band, including 5.8 and 8.0
$\mu$m bands, are given the hightest grade (=3). The lowest grade (=1) is given
to sources lacking one or several bands at $\lambda \le$4.5$\mu$m, either
because of field coverage or because of confusion problems. 

\item The third one ($Q3$) is the UV luminosity of the candidate at the best-fit
photometric redshift, after correction for lensing effects. The hightest (=3) and
the lowest (=1) grades are given respectively to sources fainter than 3$L^{*}$
and brighter than $\sim$10$L^{*}$, where $L^{*}$ stands for the 
Reddy \& Steidel\ (2009) value assuming no evolution. 

\end{itemize}

   As a result of these criteria, the most likely sources are given the highest
cumulated value of $Q=Q1+Q2+Q3$, allowing us to define a final grade which
represents the quality of a given candidate, ranging between 3 and 9 (for an
ideal candidate). As seen in Table~\ref{tab_catalog1}, four candidates achieve
the highest rates, between $Q=$7 (z1) and $Q=$8 (Y3, Y7 and J1). In all these
cases, the high intrinsic luminosity is responsible for a lower value of
Q3 with respect to the ideal case. These sources are considered as highest
quality or category I. Three candidates achieve a fair value of $Q=$6 (Y4, Y5
and Y8) and are therefore considered as reasonably good (category II) candidates. 
The lowest grade (category III) is achieved for sources with close neighbours
potentially affecting the quality of the global SED (Y1, Y2 and Y6). 

  The second grade is given by the optical $\chi^2_{opt}$ computed on the
optical bands as follows (see also Bouwens et al.\ 2010):
\begin{equation}
\chi^2_{opt} = \sum_i SGN(f_i) (\frac{f_i}{\sigma_i})^2
\end{equation}
where $f_i$ is the flux in the band $i$, $\sigma_i$ is the corresponding
uncertainty, and $SGN(f_i)$ is equal to 1 if $f_i>$0 and equal to -1 if
$f_i<$0. We have used $IRAF$ package $qphot.apphot$ to measure fluxes in a
1.3'' diameter aperture, together with the corresponding noise in the
neighbouring sky region. $\chi^2_{opt}$ values reported in
Table~\ref{tab_catalog1} are based on $I$ and $z$-band images. In the case of
J1, the $Y$-band was also included in the $\chi^2_{opt}$
calculation. Straightforward simulations were conducted in order to determine
the $\chi^2_{opt}$ distribution expected for truly non-detected sources as well as
for sources at different S/N values, in particular for those close to the
2$\sigma$ non-detection criteria in $I$ and $z$. All genuine 
non-detected sources exhibit $\chi^2_{opt} <$ 2, with 90\% at $\chi^2_{opt} <$
1 level, whereas only 3\% of sources with S/N$\sim$2$\sigma$ in $I$ and $z$
are found with $\chi^2_{opt} <$ 2 (1\% with $\chi^2_{opt} <$ 1). As seen in
Table~\ref{tab_catalog1}, all our candidates exhibit $\chi^2_{opt}$ \ltapprox
1, the highest values corresponding to Y1 and Y2 (already ranked among the
category III above). 

These two quality grades above provide 
a useful priority for spectroscopic follow up,
although they do not take into account all the details regarding 
SED-fit constraints, as discussed below, which are somehow model-dependent.
In particular the 24\micron\ emission, which is difficult to reconcile with a
high-$z$ identification for z1 and J1 despite a high grade. Also the detection
of J1 on the HST $z_{850}-$band favours a low-$z$ solution for this source,
which is considered hereafter as a possible interloper.


\subsection{Photometric redshifts with a luminosity prior}
\label{photoz_prior}

Given the luminosities derived for these candidates in the high-$z$ hypothesis, 
leading to rather extreme masses and star-formation rates, it seems likely
that, despite a best-fit at high-$z$, a large fraction of them actually
corresponds to low-$z$ interlopers. In order to better quantify this
contamination, we have introduced a luminosity prior when computing
photometric redshifts.  

   A prior probability distribution was introduced as a function of redshift
and magnitude, following Benitez (2000). In this case, the prior probability
is the redshift distribution for galaxies of a given apparent magnitude $m$. 
Given the wide redshift domain covered by the $P(z)$, and the fact that we are
likely dealing with either genuine high-$z$ or $z\sim$1.5-2.5 star-forming
galaxies, we computed the prior probability based on the luminosity function
for star-forming galaxies in the $B$-band (Ilbert et al.\ 2005). This band is
indeed directly ``seen'' by the SED of galaxies in our sample for all
redshifts between $z\sim$0.8 and 9. A smooth probability distribution prior was
computed for each object as a function of redshift, with the absolute
magnitude M$_B$ derived from the apparent magnitude $m$ which is closer to the
rest-frame $B$-band. The final probability distribution is given by the
previous $Hyperz$ $P(z)$ multiplied by the prior. 

\begin{figure}[htb]
\centering{
\includegraphics[width=0.48\textwidth]{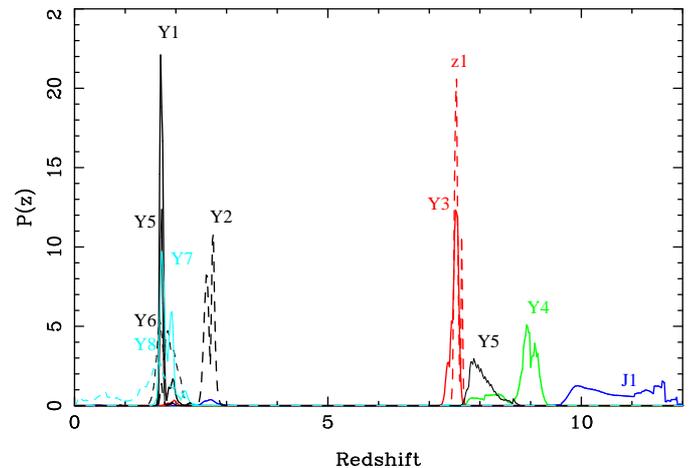}
}
 \caption{Redshift probability distributions for the sample of high-$z$
candidates, arbitrarily normalized to 100 between $z=$0 and 12, 
resulting from $Hyperz$ $P(z)$ multiplied by a smooth luminosity prior.
}
\label{fig_prior}
\end{figure}

   Fig.~\ref{fig_prior} displays the resulting probability distributions for
all candidates, arbitrarily normalized to 100 between $z=$0 and 12. As seen in
this figure, four candidates still exhibit a dominant high-$z$ solution,
namely z1, Y3, Y4, and J1, whereas one candidate is degenerate between low and
high-$z$ (Y5). The best high-$z$ candidates also exhibit the better quality
grades, as seen in previous section. We use these results to propose a
final tentative classification between likely low-$z$ interlopers and high-$z$
candidates in Table~\ref{tab_catalog1}. Despite a high grade, z1 and J1
are ranked among the likely low-$z$ contaminants based on the 
detections in the 24$\mu$m and $z_{850}$ bands respectively. 

\subsection{Physical properties from SED fits}
\label{properties}

\begin{figure}[htb]
\centering{
\includegraphics[width=0.48\textwidth]{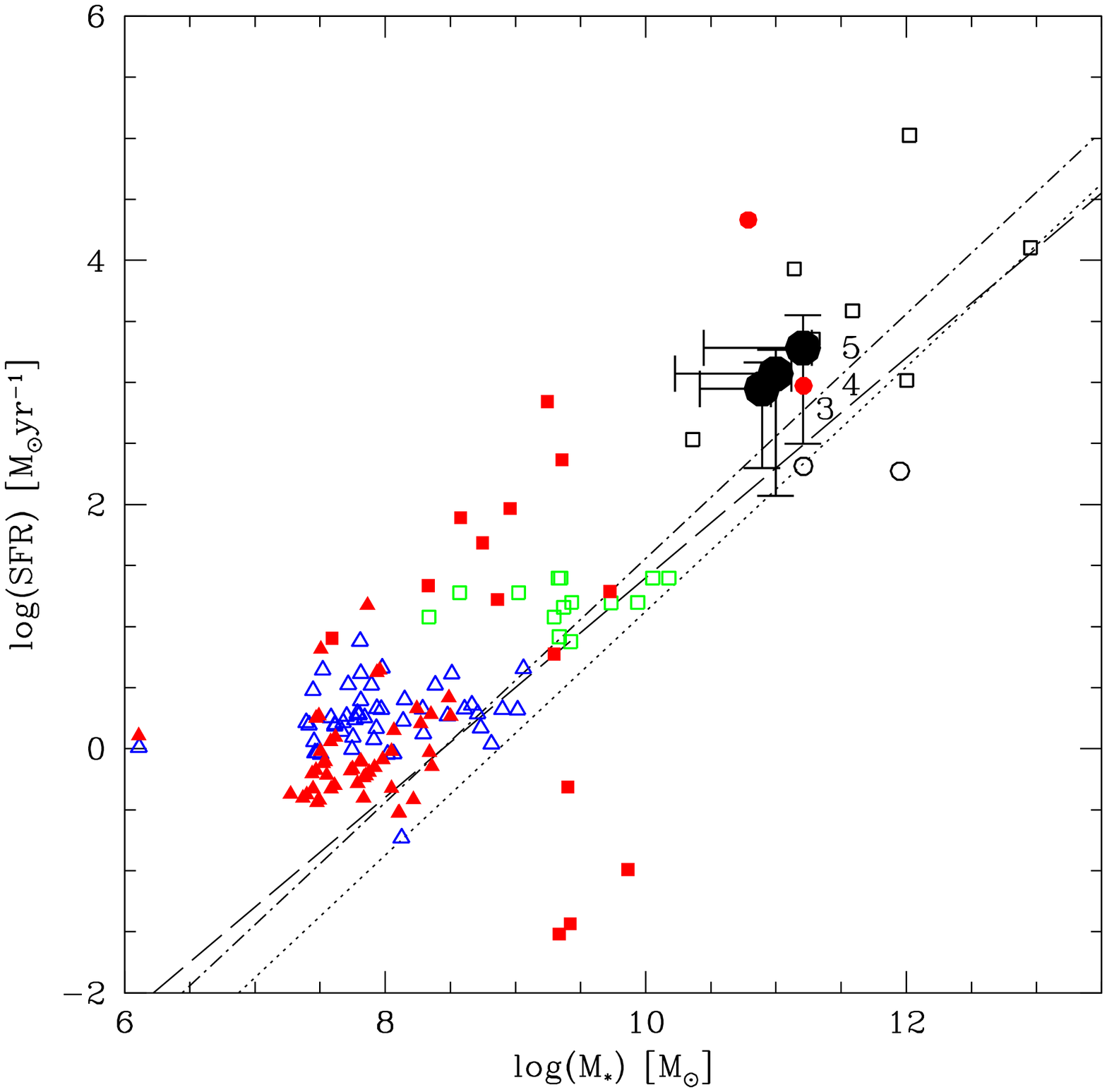}
}
 \caption{Mass--SFR relation of the $z$, $Y$, and $J$-dropout galaxies 
resulting from $10^3$ Monte Carlo simulations compared to the properties
of $z \sim$ 6--8 galaxies from recent surveys analysed by Schaerer \& de
Barros (2010).
The most likely high-$z$ candidates based on photometric redshifts
with a luminosity prior, namely 
Y3, Y4 and Y5, are identified and displayed by large black dots. The positions
derived for the other optical-dropouts in this study, if at high-$z$, are
indicated by black open squares for comparison. 
Open symbols (squares, triangles) show the ``standard'' SFR(UV) value
(not corrected for extinction) versus mass derived from the SB2010
reference model for objects from their bright, intermediate, 
and faint samples respectively. 
Red filled symbols show the best-fit model 
SFR and $M_\star $ values when assuming $\tau \ge 10$ Myr (model 1 in
SB2010). Circles correspond to the 2 objects from Capak et al.\ 2011).
The dotted (dash-dotted) lines show the locus for ${\it SFR}={\rm const.}$
from $z=\infty $ (10) to 7 corresponding to ${\it SSFR} = 1.3$ (3.6) Gyr-1. 
The dashed line shows the relation found by Daddi et al. (2007) for $z \sim 2$
star-forming galaxies. 
The large spread in SFR is in particular due to the fact that a wide variety of 
exponentially decreasing star-formation histories are allowed.
Note that if at high redshift the properties of our galaxies follow the trends
observed for less massive/bright objects a high redshift.}
\label{fig_sfr}
\end{figure}

\begin{figure}[htb]
\centering{
\includegraphics[width=0.48\textwidth]{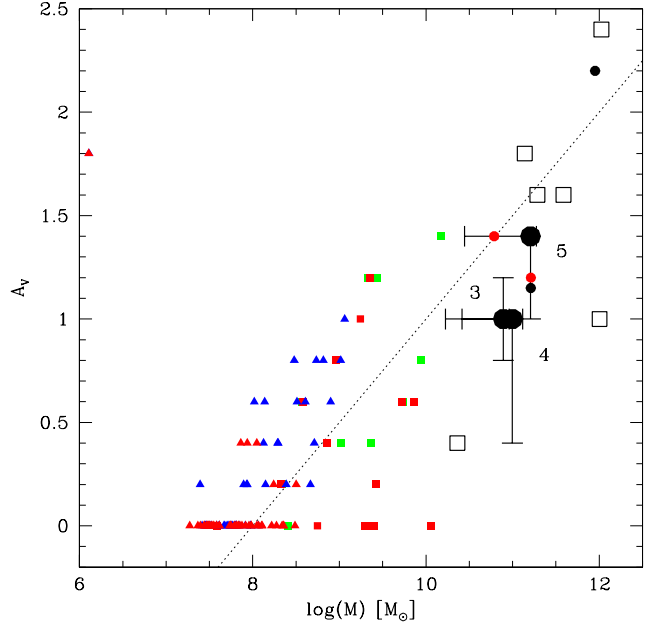}
}
 \caption{Mass--AV relation for the same objects as shown in Fig.\ \protect\ref{fig_sfr}
(same symbols).}
\label{fig_av}
\end{figure}

From our SED fits using the Bruzual \& Charlot templates we can also
derive the physical properties of the galaxies, such as the age of the stellar population,
the stellar mass, star-formation rate, and attenuation. The resulting masses, SFR, and
attenuation $A_V$ (derived assuming the Calzetti law) and the uncertainties, derived
from 1000 Monte Carlo simulations of each object, are shown in Figs.\ \ref{fig_sfr}
and \ref{fig_av}. For comparison we also show the properties of $z \sim$ 6--8 galaxies
analysed recently by 
SB2010 using the same SED fitting tool.

As can be seen, 
the masses derived for the most likely high-$z$ candidates based on
photometric redshifts with a luminosity prior, namely
Y3, Y4 and Y5, are among the highest masses found by SB2010, 
typically of the order of $M_\star \sim 2\times 10^{10}$ to $2\times 10^{11}$ \msol.
Including relatively large corrections for attenuation ($A_V \sim$ 0.4--1.4),
their SFR range from $\sim$ 100 to $\sim 10^3$ \msunyr. 
For comparison, the SFR derived from the rest-frame UV luminosity L$_{1500}$
using the Kennicutt (1998) calibration is typically $\sim$ 100 \msunyr for
these sources (see Table~\ref{tab_prop}).

When including the full sample of optical-dropout candidates, the masses and
SFRs achieved are much larger, reaching $M_\star\sim 10^{12}$ \msol in most
cases, and even $M_\star> 10^{12}$ \msol for the brightest $Y$-drop candidate,
Y1, if at high-$z$.   
However, for the most extreme objects the uncertainties on $A_V$ and SFR are
very large. For objects with well defined errors, the SFR may reach
up to 2000-3000 \msunyr.
Interestingly enough these values are not in disagreement with the trends 
found by {\bf SB2010} for the fainter $z >6$ galaxies from 
recent HST surveys, if extrapolated to higher masses. These properties are also 
similar to those of the two
$z \sim 7$ galaxy candidates found by Capak et al.\ (2011) in the COSMOS wide
field survey. However, whether these relatively bright objects are truely high
redshift galaxies, and hence objects with such extreme properties, remains
of course questionable (see below). 

\begin{figure*}[p]
   \centering
\includegraphics[angle=270,width=1.0\textwidth]{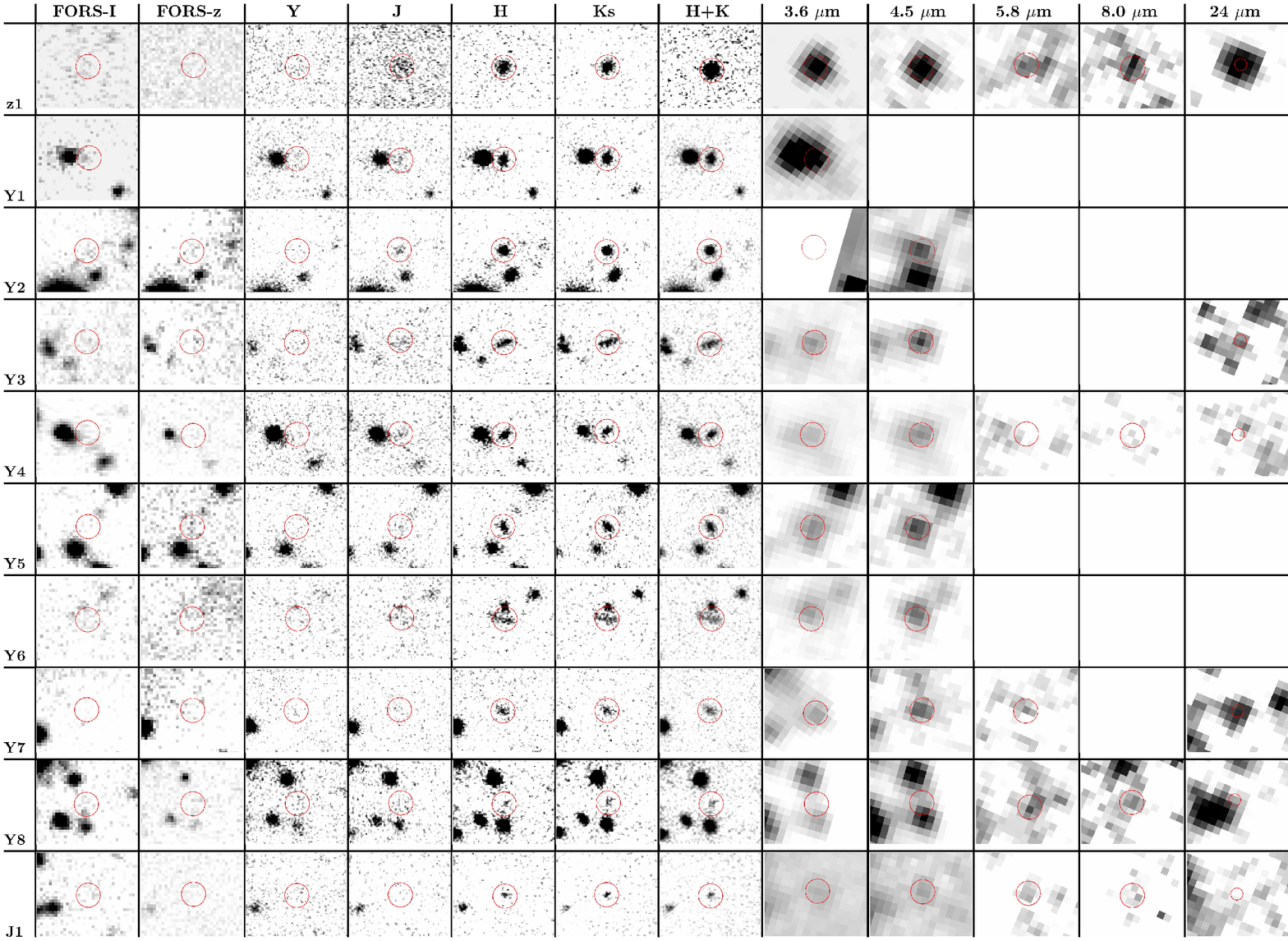}
   \caption{Thumbnail images of the bright $z$, $Y$ and $J$-dropouts found in
A2667, covering a $9'' \times 7.3''$ area on the original images from $I$ band
to 8.0$\mu$m, and $18'' \times 14.6''$ for MIPS 24$\mu$m.  
The position of each candidate is displayed by a circle of 2\arcsec
diameter aperture. Images are displayed in linear scale ranging between -5 and
20 $\sigma$ around the sky background. 
}
              \label{trombino}%
    \end{figure*}

 
   Two of our optical dropout galaxies (z1 and Y7) are detected by MIPS at 24
\micron\ with fluxes 3.4 and 1.1 $\times 10^{-4}$ Jy respectively, and we have 
non-detection constraints for three additional sources included in the MIPS
image (Y3, Y4 and J1, with 1$\sigma$ fluxes below 38.7 $\mu$Jy), whereas Y8
is highly contaminated by neighbouring galaxies (see Fig.~\ref{trombino}). 
Recently, two objects of this sample have also been detected with Herschel and
LABOCA between 160 and 870 $\mu$m (Boone et al., in preparation). This data
identifies z1 and Y5 as mid-$z$ interlopers.

\begin{figure*}
\centering{
\includegraphics[width=0.42\textwidth]{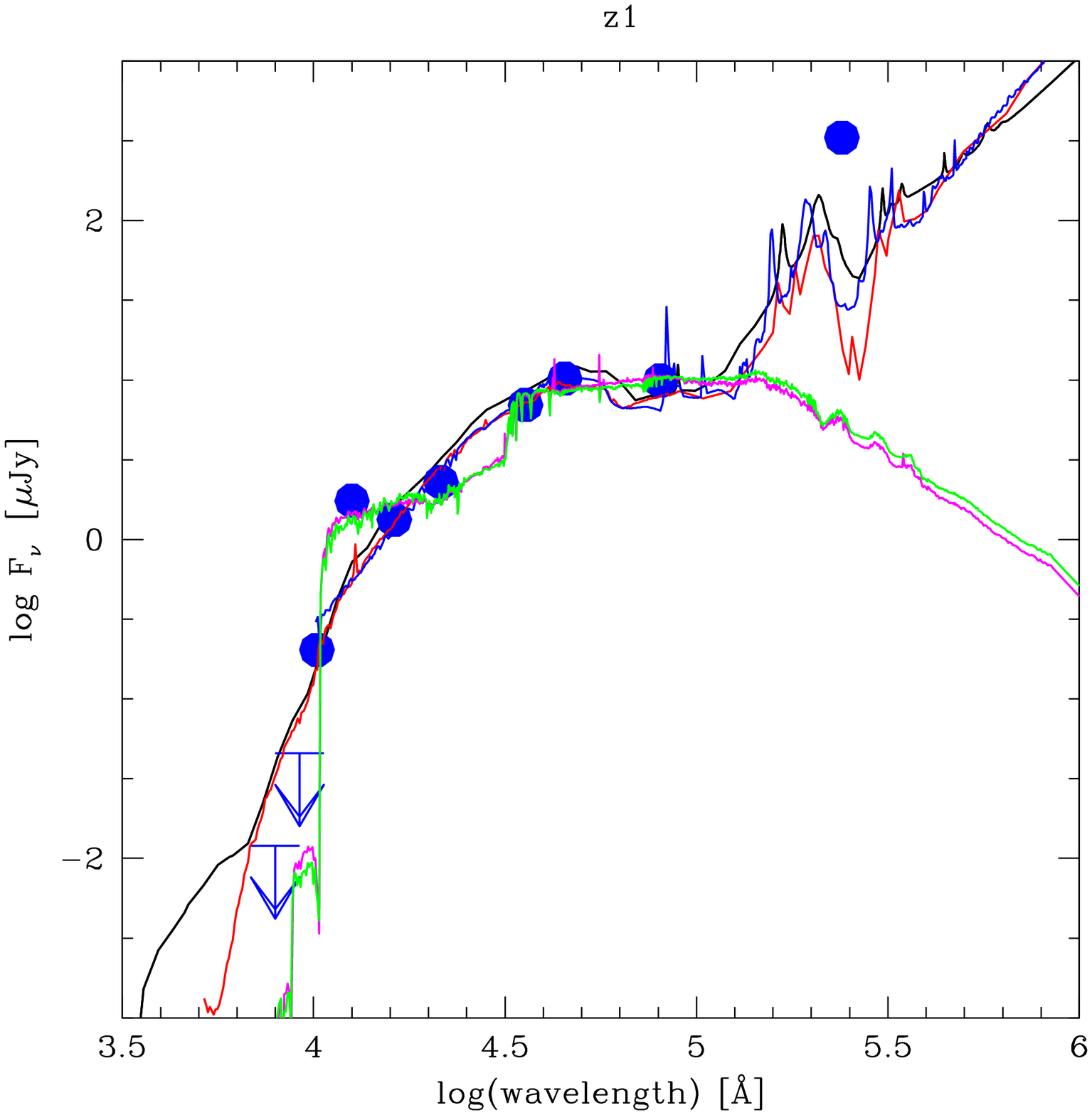}
\includegraphics[width=0.42\textwidth]{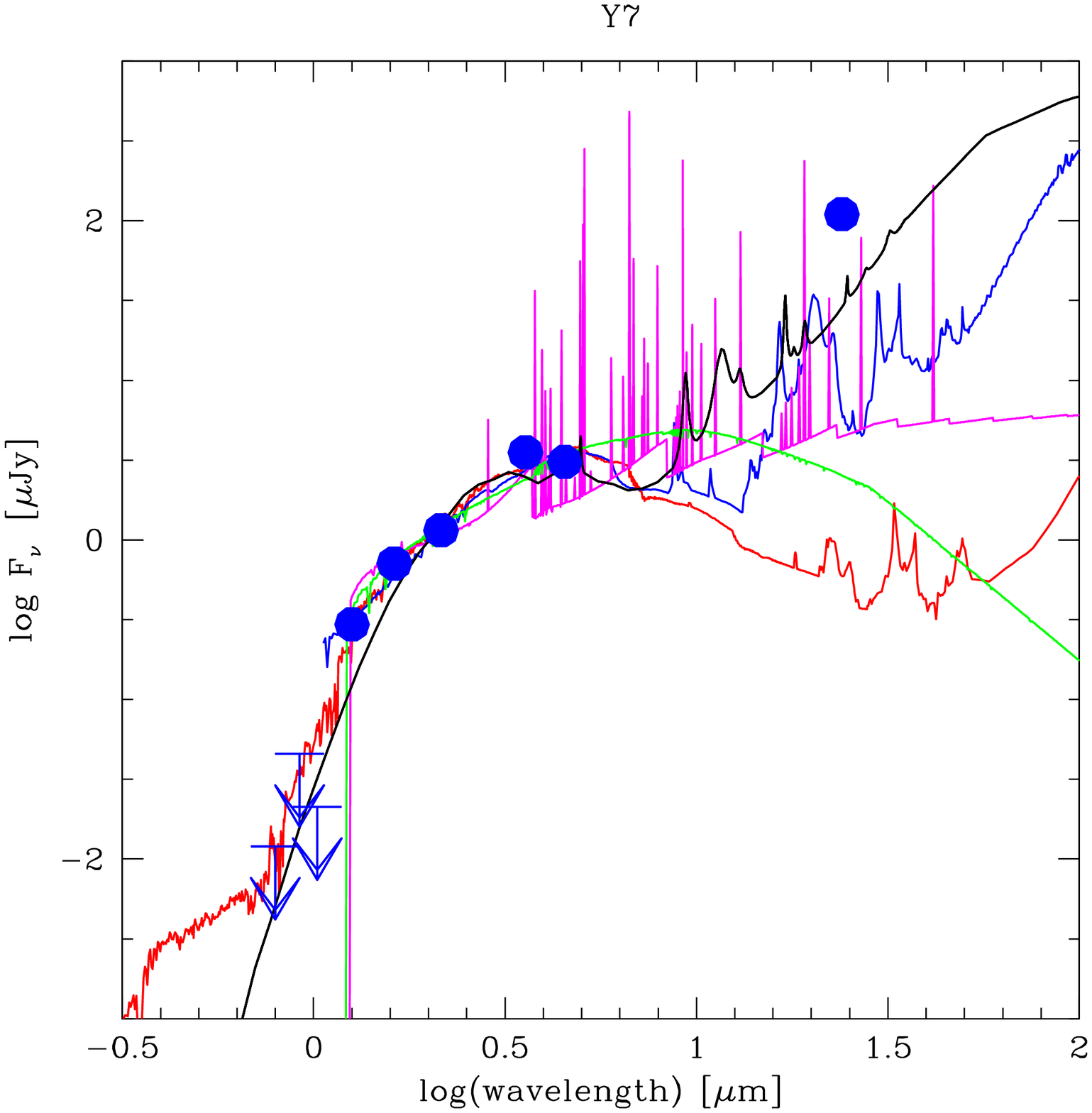}
}
\centering{
\includegraphics[width=0.42\textwidth]{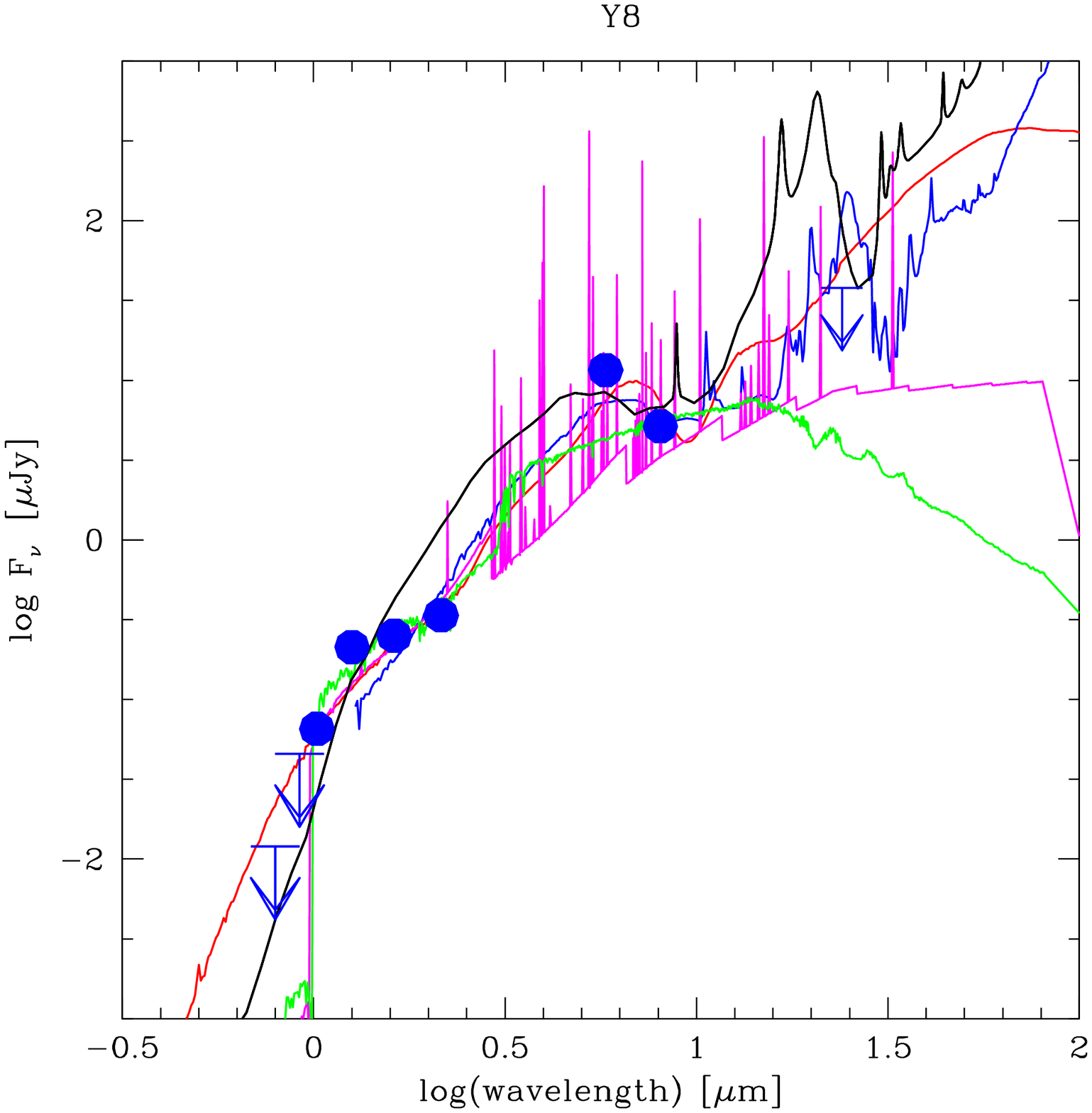}
\includegraphics[width=0.42\textwidth]{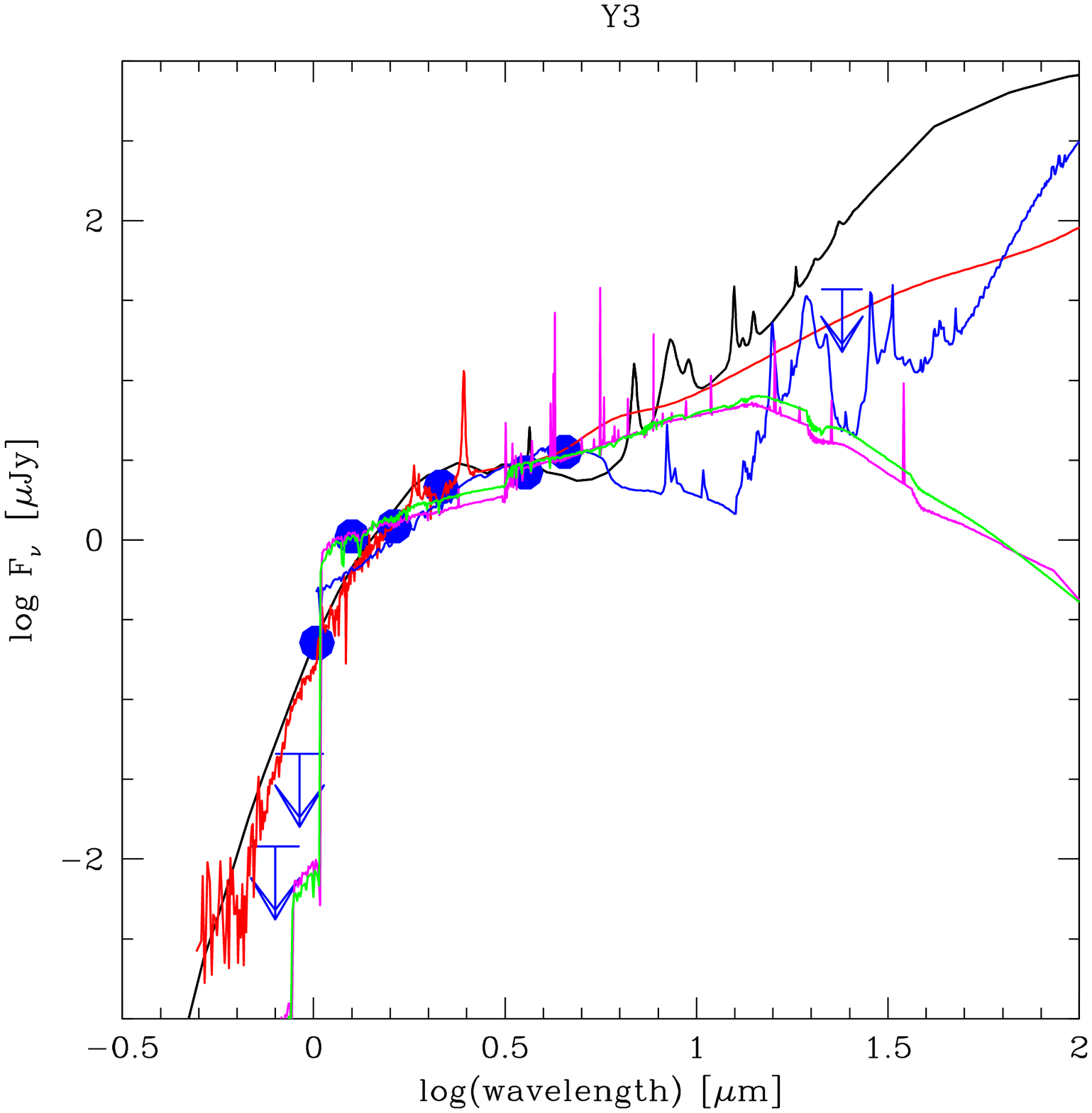}
}
\centering{
\includegraphics[width=0.42\textwidth]{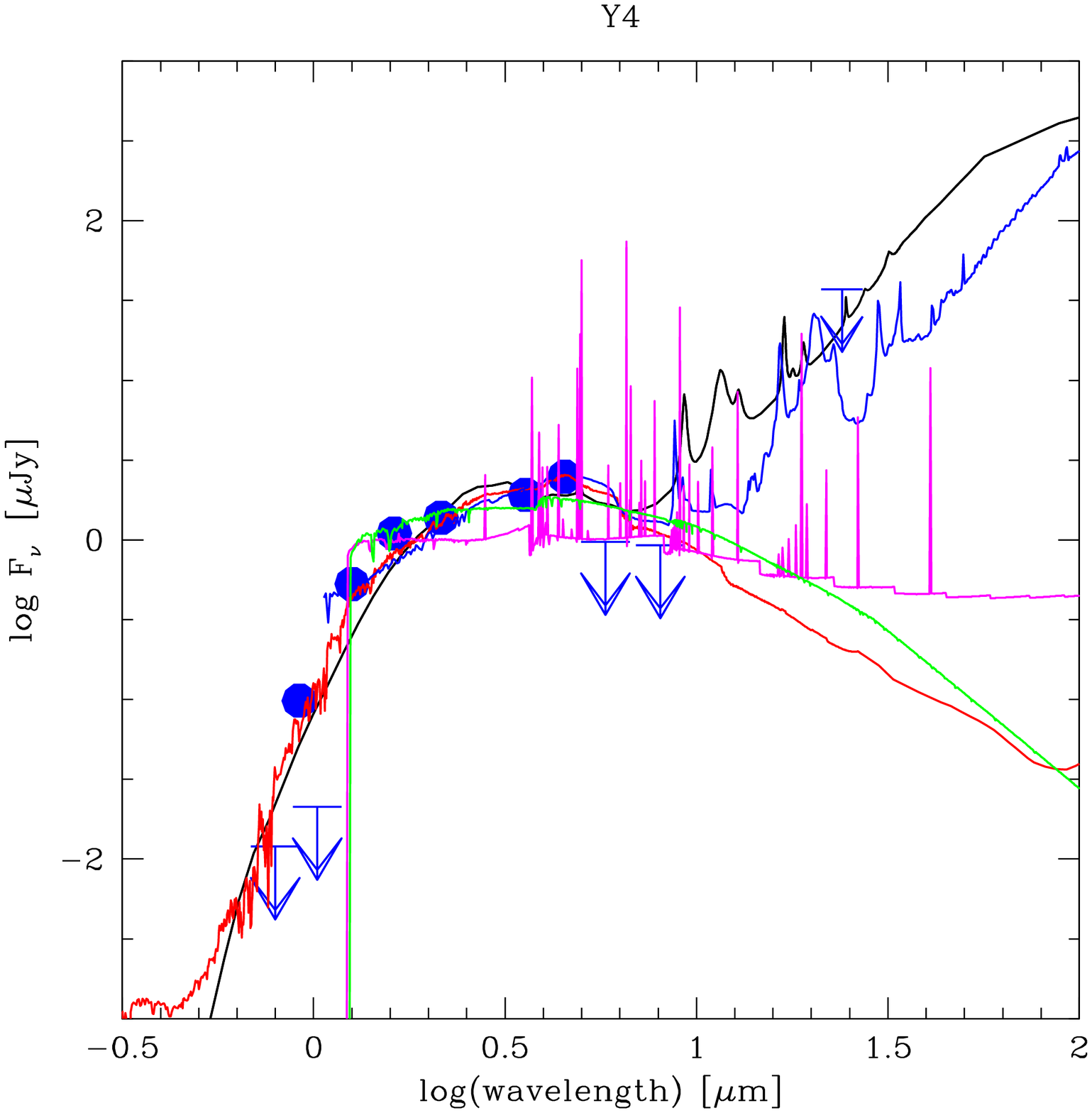}
\includegraphics[width=0.42\textwidth]{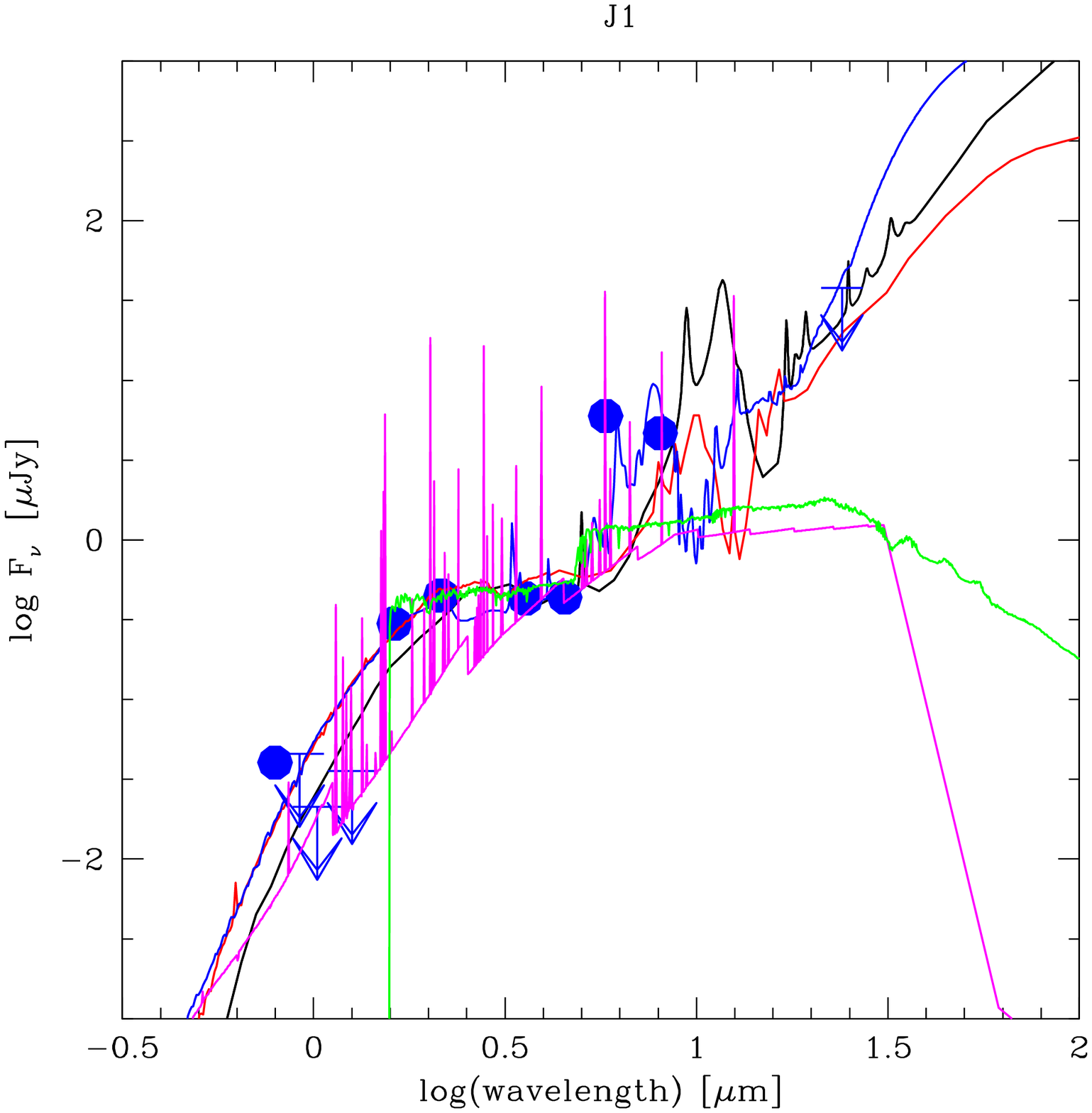}
}
 \caption{Best-fit SEDs for candidates with available constraints at 24
\micron\ : z1 and Y7 (sources clearly detected, top row),
Y8 (source highly contaminated, middle left panel), and 
Y3, Y4 and J1 (non-detection, 1$\sigma$ flux $<$ 38.7 $\mu$Jy). 
Different spectral templates are displayed.
Green (pink) lines show the best-fit solutions at $z>7$ using our standard
templates based on Bruzual \& Charlot models without (with) nebular lines.
Black and red lines show the best-fits at low redshift ($z<4$) using the 
templates from the GRASIL models (Silva et al.\ 1998) and the 
SWIRE starburst-AGN templates of Polletta et al.\ (2008) respectively.
Blue lines show best-fits using the empirical LIRG--ULIRG templates 
of Rieke et al.\ (2009).
}
\label{fig_mips}
\end{figure*}

For  $z \sim$ 1--2 galaxies the MIPS band probes a region in the mid-IR
corresponding or close to redshifted PAH emission. If at high redshift
($z \sim$ 7--9), the 24 \micron\ band samples the region between 2--3 \micron.
To illustrate the overall SED of our MIPS-detected (or constrained) objects we
show their photometry together with several SED fits in Fig.\ \ref{fig_mips}.
In addition to our standard spectral templates with/without nebular emission
we also show best-fits using (semi)empirical templates of nearby galaxies
including in particular very dusty galaxies, LIRG, and ULIRG.
In practice we have used the templates from the GRASIL models of Silva et al.\ (1998),
the  SWIRE starburst-AGN templates of Polletta et al.\ (2008), and 
LIRG--ULIRG templates of Rieke et al.\ (2009); redshift and additional extinction 
are kept as free fit parameters. For all of these models we here show the best-fits
under the constraint that $z<4$ and without any constraint. 
We note that the best-fit photometric redshifts at low-$z$
can differ quite significantly depending on the set of spectral templates used and between
the objects. For example, we find \zphot $\sim$ 0.1, 1.5, and 2.7 for Y3; 1.6--1.7 for Y7, 
0.02, 1.6, and 1.6 for Y8, and 0.1--0.5 for J1.
Furthermore the 24 \micron\ flux expected from these fits can vary
by more than an order of magnitude. Interestingly fits at high-$z$ show fluxes
at 24 \micron\ which are comparable to those expected from some of low-$z$
best-fits, thus providing weak additional constraints on the redshift. 

   Several objects in this sample deserve additional specific comments.
\begin{itemize}

\item z1: This is the brightest source in our sample. 
It is an extended isolated object, clearly non-stellar. It is detected in
all filter bands between $Y$ and 24\micron\, except at 5.8\micron\ where
no data is available. Although the best-fit is obtained at high-$z$, 
even with a luminosity prior, 
the 24 \micron\ flux seems incompatible with the high-$z$ solution (see
Fig.~\ref{fig_mips}). It appears even too bright in this band when compared to the
best-fit templates at low-$z$. 
However, the observed flux in the $J-$band, and the fact
that, if at low-$z$, it should be also detected in the $I$ and $z$ bands at
$\ge$ 3 $\sigma$ level (given the depth of our survey), remain difficult to
explain.

\item Y3: Although the 24 \micron\ non-detection of this object helps to exclude 
at least one of the GRASIL templates at low-$z$, it is still consistent with
other templates at low and high-$z$. Probably the main failure of the low-$z$
templates is again the mismatch in $J$ and in the optical bands. 

\item Y4: As for Y3, the upper limit detection at 24 \micron\ provides
weak additional constraints on the redshift. The non-detection at 5.8 and
8\micron\ are interesting because both solutions at low and high-$z$ seem to
predict a detection at $\sim$2-3$\sigma$ level. 
The object is formally detected at the \ltapprox$2 \sigma$ level in the $z_1$ band, but
it is not detected in the  $z_2$ image. 

\item Y7: For this object, the expected 24 \micron\ fluxes for the
best-fit at high and low redshifts are very similar. In particular, the 
flux of the $z \sim 9$ fit is even quite similar to the brightest flux
predicted from dusty galaxy templates at $z \sim 1.6$. Therefore, this object
cannot be excluded as a genuine high-$z$ based on its MIPS photometry
without introducing a luminosity prior.
Besides, the observed 24\micron\ flux seems too high irrespective of the
template redshift for a relatively isolated source (see Fig.~\ref{trombino}). 

\item Y8: The 5.8 and 8\micron\ photometry for this object could be
seriously contaminated by bright neighboring sources (see Fig.~\ref{trombino}),
although this seems to be the only emission at this precise location. The
24 \micron\ flux shown in Fig.~\ref{fig_mips} corresponds to the lower limit,
and it is compatible with both the low and the high-$z$ solutions. 

\item J1: This is the faintest candidate in our sample, and the only
one which is located on the central area covered by the HST. 
The best-fit solution is found at high-$z$ 
even when including the $z_{850}-$band flux and a luminosity prior,
as seen in Table~\ref{tab_prop}. However, the detection on the $z_{850}-$band
makes the high-$z$ solution unlikely.
The upper limit detection at 24
\micron\ provides weak additional constraints on the redshift. The detection
on the HST image is fully consistent with the large break identified on
ground-based images. 
The apparent strong ``double-break'' between the optical/near-IR and between
4.5/5.8 \micron\ seems quite unusual. 
High-$z$ solutions have difficulty reproducing the latter break (see
Fig.\ \ref{SED-fit}), while low-$z$ solutions predict a flux excess
in the optical domain. 
The break at 4.5--5.8 \micron\ could be explained by PAH features boosting
the 5.8 and 8 \micron\ fluxes, as shown in Fig.~\ref{SED-fit}. This is one of
the candidates for which a spectroscopic follow up is needed to conclude. 

\end{itemize}




\subsection{Contamination level based on stacked images}
\label{stacked}

Based on previous discussions, a large fraction of optical-dropout
galaxies in this sample could be low-$z$ contaminants, reaching 
as high as $\sim$70\% based on luminosity priors. In order to better quantify
this estimate, we have generated stacked images in the $I$ and $z$ bands,
where genuine high-$z$ galaxies are not expected to be detected. On the
contrary, the S/N ratio achieved on the stacked image should allow us to
estimate the contamination level. 
   
    For each candidate, a 10\arcsec $\times$ 10\arcsec region has been
selected in the $I$ and $z$-band images around the centroid position 
on the detection image $H+Ks$. An additive zero level correction has been
applied to each single image to properly remove the local average sky
background. $I$ and $z$-band images have been averaged using IRAF routines and
different pixel rejection schemes in order to improve the suppression of
neighboring sources on the final combined images. Due to the presence of a
closeby galaxy, we do not include the Y1 field in the final stacks. We have
checked that the sky backgound noise on the final combined images actually
improves as expected as a function of the total number of images stacked,
i.e. reaching a reduction by a factor close to 3 on the final stacks with
respect to the original images. 
 
    We used IRAF routines to measure fluxes and magnitudes in a 1.3'' diameter
aperture, together with the corresponding error bars on the different stacks. 
The best S/N ratio is achieved in the $I$-band, where we measure up to 
$I=$28.0$\pm$0.3 for the combined source, i.e. a S/N$\sim$3. The detection in
the $z$-band is less significant, reaching $z=$27.7$\pm$0.5 (S/N$\sim$2) on the 
best-detected final stack. These results confirm that there is indeed some
contamination by low-$z$ interlopers in our sample. The detection level in the
$I$-band is roughly consistent with $\sim$70\% of the sample being detected
at $\sim$1$\sigma$ level, although 3-4 objects detected between 1 and 2
$\sigma$ would be enough to account for the signal in this band. The flux
measured in the $z$-band seems to favour either a signal below
1$\sigma$ for a large majority of our candidates, or a higher signal coming
from a small fraction of contaminants (between $\sim$20\% with S/N$\la$2, or
$\sim$30\% with a mean S/N$\sim$1.5). 

\section{Discussion}
\label{discussion}

   As the selection diagrams (Figs.\ \ref{CC_YJH} and \ref{CC_JHK}) show, most
of our bright candidates have near-IR colors distinguishing them clearly from
normal galaxies at low redshift and late type stars. Also SED-fitting results clearly
favor a high-$z$ solution for all these candidates, irrespective of
their intrinsic luminosities. However, as seen in Table~\ref{tab_prop},
high-$z$ solutions yield magnification-corrected luminosities which typically
range between 3 times (Y8) and 40 times (z1, Y2) L$_{1500}^*$ at these
redshifts, according to the evolving LF by Bouwens et al.\ (2008), suggesting
a potential contamination by low-$z$ interlopers. In this 
section we discuss the observed versus expected number counts of high-$z$
galaxies, and the possible sources of contamination in our survey. We also
compare our results and the properties of this sample with those obtained by
previous authors using similar techniques.  

\subsection{Observed versus expected number density of high-$z$ sources}
\label{comparison_models}

   We first compute the expected number counts of bright high-$z$ sources in
this lensing field, as a function of redshift and for the range of  
magnitudes of our candidates (i.e. AB$\sim$23 to 25.5), and we compare these
numbers with current observations. This calculation was done following the
same procedure as in Maizy et al.\ (2010). All the noisy regions 
in the field, in particular around bright galaxies in the cluster core, have
been masked.
The presence of a strong lensing cluster in this field introduces two opposite
effects on the number counts as compared to blank fields. Gravitational 
magnification increases the number of faint sources by improving
the detection towards the faint end of the LF, whereas the dilution effect
reduces the effective volume by the same factor. As discussed in Maizy et
al.\ (2010), the difference between lensing and blank field results 
depends strongly on the shape of the LF. We expect lensing clusters to be more
efficient than blank fields in relatively shallow surveys. 

   Number counts of sources brighter than a limiting magnitude were computed
by a pixel-to-pixel integration of the magnified source plane as a function of
redshift (see eq. 6 in Maizy et al.\ 2010), for redshift bins $\Delta z=1$,
using the evolving LF by Bouwens et al.\ (2008), i.e. with the Schechter
parameters directly derived from eq. 3 in their paper. 
For comparison purposes, we also derive the expected counts with
the Beckwith et al.\ (2006) LF for $z\sim6$ galaxies, assuming no
evolution. This LF displays the same Schechter parameters as for Steidel et
al.\ (2003) and Reddy \& Steidel\ (2009),
but the normalization factor is 3 times smaller than for Steidel et
al.\ (2003).   
Table~\ref{N_counts} summarizes these
results for two different limiting magnitudes, AB$<$25.5, where all the
current candidates are found, and AB$<$26.0, which corresponds to our
$\sim$5$\sigma$ detection level (in the filter spanning the UV region around
1500\AA\ ). Error bars include Poisson uncertainty and field-to-field
variance, following Trenti \& Stiavelli\ (2008). Given the small
number counts expected, these fluctuations dominate the error budget (see also
the discussion in Maizy et al.\ 2010). We also computed number
counts using the latest Schechter parameters presented by Bouwens et al.\ (2010)
for galaxies at $z\sim$7 in the $z=[7.0,8.0]$ interval, and for galaxies at 
$z\sim$8 in the $z=[8.0,9.0]$ interval (identified by $*$ in
Table~\ref{N_counts}). The changes with respect to the 2008 version are
relatively minor in this case given the luminosity domain, which is largely
dominated by statistical uncertainties. 

\begin{table}
\begin{center}
\caption{\label{N_counts}
Number counts of sources with AB$<$25.5 and AB$<$26.0 
expected in the field of A2667, with and
without the presence of a lensing cluster, for different redshift bins and for 
two different LFs: Bouwens et al.\ 2008 (1) and Beckwith et al.\ 2006 (2) (see
text). 
Error bars in number counts, including Poisson uncertainty and field-to-field
variance, are given in brackets.  
}
\begin{tabular}{l|cc|cc}
\hline \hline
           &   LF(1) &  &  LF(2) &    \\
Redshift   & Blank & A2667 & Blank & A2667 \\
\hline
AB$<$25.5  &  &  &  & \\
\hline
$[6.0,7.0]$  & 0.5 (1) & 1.5 (2) & 2.9 (3) & 4.2 (3) \\
$[7.0,8.0]$  & $<$0.1 (1) & 0.3 (1) & 1.5 (2) & 2.3 (2) \\
 (*)         & 0.1 (1)    & 0.5 (1) &         &         \\
$[8.0,9.0]$  & $<$0.1 (1) & 0.1 (1) & 0.8 (1) & 1.4 (2) \\
 (*)         & $<$0.1 (1) & 0.2 (1) &         &         \\
$[9.0,11.0]$ & $<$0.1 (1) & 0.1 (1) & 0.6 (1) & 1.3 (2) \\
\hline
AB$<$26.0  &  &  &  & \\
\hline
$[6.0,7.0]$  & 3.1 (3) & 5.4 (4) & 8.5 (5) & 10.5 (5) \\
$[7.0,8.0]$  & 0.4 (1) & 1.1 (2) & 4.9 (3) & 6.4 (4) \\
 (*)         & 0.8 (1) & 1.9 (2) &         &         \\
$[8.0,9.0]$  & $<$0.1 (1) & 0.3 (1) & 2.9 (3) & 4.0 (3) \\
 (*)         & 0.3 (1) & 0.7 (1) &         &         \\
$[9.0,11.0]$ & $<$0.1 (1) & 0.1 (1) & 2.8 (2) & 4.3 (3) \\
\hline
\end{tabular}
\end{center}

\end{table}

  Candidates were selected on a total
survey (clean) area of $\sim$42{\rm arcmin}$^2$
($\sim$33{\rm arcmin}$^2$ when corrected for dilution), i.e. an effective
lensing-corrected comoving volume per unit $\Delta z=1$ ranging between 7.4
and 4.9 $\times 10^{4}$ Mpc$^{-3}$ (covolume). 
For the comparison with the current sample, 
we consider four redshift bins, the same presented in Table~\ref{N_counts}, 
and we bin the candidates as follows
(cf.\ Table \ref{tab_prop}):
z1, Y1, Y3, Y6, Y8 (i.e.\ 5 objects) in bin $[7.0,8.0]$, 
Y2, Y5 (2 objects) in bin $[8.0,9.0]$, and 
Y4, Y7 (2 objects) in bin $[9.0,11.0]$. J1 is also in this last bin, but the
high-$z$ hypothesis is highly unlikely in this case. 

    Assuming the evolving LF by Bouwens et al.\ (2008) or Bouwens et
al.\ (2010), we expect up to a
maximum of 1-2 sources at $z \sim 7.0-8.0$ in this wide field with AB$<$25.5,
and typically between 2 and 10 sources with the Reddy \& Steidel\ (2009) LF, all
of them within 24.5$<$AB$<$25.5. In our sample, only 2 out of the 5 $z
\sim7.5$ candidates are included in this magnitude interval (Y6 and Y8), 
in full agreement with expectations for an evolving LF, whereas z1,
Y1 and Y3 seem too bright. On the other hand, only Y3 qualifies as a high-$z$
candidate when using a luminosity prior (see Sect.~\ref{photoz_prior}). This
result is also consistent with the expectations for a Beckwith et al.\ (2006)
LF. 

    Regarding the number of bright (24.5$<$AB$<$25.5) higher-redshift sources
expected at $z \approx 8.5$ and $z \approx 10$ in this field, it ranges
between a maximum of one per bin for an evolving LF, and typically between 1--7
per bin with a constant $z\sim3$ LF, when including the error
bars. Only Y7 and J1 seem to be in agreement with both redshift interval and
observed magnitude, whereas Y2, Y4 and Y5 seem too bright when taken at face
values to be all at such high redshifts, i.e.\ we have less than 1/6 chances
to find one such an intrinsically bright object in this field, even when
assuming no evolution in the LF since $z \sim 4$. On the other hand, only Y4
and Y5 qualify as high-$z$ candidates when using a luminosity prior to derive
photometric redshifts (see Sect.~\ref{photoz_prior}), 
i.e. a maximum of two candidates per bin. The result is the same when blindly 
excluding the brightest candidates,
as well as J1 for arguments related to its SED (cf.\ above), our results seem
to be in agreement with an evolving LF, and also consistent with Beckwith et
al.\ (2006) counts at $z\sim6$ when error bars in number counts are taken into
account.  

    In summary, our sample includes some intrinsically bright 
sources (6 out of 10) for which the best-fit photometric redshifts seem
difficult to reconcile with the LF previously measured at high-$z$, even 
when no evolution is assumed. When these sources are blindly excluded,
however, 
or when the sample is restricted to galaxies surviving a stringent
prior in luminosity (namely Y3, Y4 and Y5), 
observed number counts at z$\gtrsim 7.5$ are in agreement with
expectations for an evolving LF, also consistent within the error bars with a
constant LF since $z \sim 6$, and inconsistent with a constant LF
since $z \sim 4$.


\subsection{Contamination}
\label{contamination}

   As seen in Sect.~\ref{comparison_models} based on pure LF and counts
arguments, 6 out of 10 candidates, seem too bright (or their abundance is too
high) as compared to expectations. 
Also half of our sample could be identified as mid-$z$ interlopers when
computing photometric redshifts including a luminosity prior, and two galaxies
among the surviving sample can be excluded based on their SED properties (z1
and J1), leading to a contamination level close to 70\%.
We discuss below on
the possible sources of contamination, which must be also present in other
current Lyman Break surveys.

    Regarding the selection windows, several objects are relatively close to
the boundaries. Object Y8, which also shows a high stellarity index, has
colors indistinguishable within the errors from late-type stars (cf.\ also
Fig.\ \ref{CC_ZJK}). However, although its flux could not be properly
extracted due to blending, it seems to be detected at 24 \micron, which
excludes a Galactic late-type star. Most likely, it is therefore a dusty, low
redshift galaxy, cf.\ Fig.\ \ref{fig_mips}. 
Objects Y3 and Y6 are also relatively close to the boundaries of the Y-drop
selection box, where stars and low-redshift galaxies lie.
However, when colors spanning a wider wavelength range are considered, as in Fig.\
\ref{CC_ZJK}, the difference from ``contaminants'' is more pronounced, especially 
for Y3. 

   As Fig.\ \ref{CC_ZJK} shows, 4 out of 9 sources (z1, Y2, Y3, Y5) with
$z$-band photometry show colors (z-J) $\ga$ 3, redder than the most extreme
low redshift objects compiled by Capak et al.\ from the large field COSMOS
survey. Six of our 10 objects (z1 and Y1-Y5) also show (I-J) $>4$, i.e.\ the depth
needed in the z-band to robustly select $z>7$ galaxies, according to
the comparison with COSMOS galaxies (cf.\ Capak et al.\ 2011).
In addition, our data also comprises the $Y-$band, yielding a stronger constraint
on the shape of the SED than available e.g.\ for the COSMOS sample.

   In other color-color diagrams, such as (J-K) versus (J-3.6) examined by
Stanway et al.\ (2008) and Capak et al.\ (2011), our objects are where 
expected for high-$z$ objects, and they show similar colors as the two
$z>7$ candidates of Capak et al.\ (2011). The (3.6-4.5) color is also
as expected from these papers. 

   Could emission line objects, such as the ultra-strong emission-line
galaxies (USELs) recently discovered by Kakazu et al.\ (2007) at $z \sim$
0.3--1.5 contaminate our sample? In principle, their strong O~[{\sc iii}]
$\lambda\lambda$4959,5007, H$\alpha$ and other emission lines could lead to
blue (J-H) colors and a spectral break if at $z\sim$ 1.6--1.8, 
similar to the behaviour expected for high-$z$ galaxies.
However, from the objects properties of the USELs known so far this seems quite
unlikely, for several reasons. If we assume a rest-frame equivalent width of $\sim$ 1000 \AA\
for both  O~[{\sc iii}] $\lambda$5007, and H$\alpha$, as observed for the most extreme
objects (cf.\ Kakazu et al.\ 2007), their contribution to the broad band $J$ and $H$ filters
(with widths of $\Delta \lambda \sim$ 1400--2700 \AA) should at best be 
$\sim$ 35 --50 \%. Assuming a flat underlying continuum (in $F_\nu$),
as expected for strongly star-forming objects with little/no extinction
and roughly also consistent with their observed colors 
(cf.\ Kakazu et al.\ 2007, Hu et al.\ 2009), these emission
lines can therefore not mimick a spectral break much larger than $\sim$ 0.75
mag. Hence such objects, if existant at $z \sim$ 1.6--1.8 are most
likely unable to reproduce the break of $(Y-J) \gg 1$ shown by most of our
objects. 

Our spectral models, allowing also the presence of nebular emission (lines
and continua) find indeed some extreme best-fit templates when $z<4$ is imposed.
This is for example the case for Y8 and J1, where our fitting procedure exploring
a wide range of parameter space identifies relatively young ($< 10$ Myr) objects
with a very strong extinction ($A_V \sim$ 3--3.8) as the best-fits at low redshift,
as shown in Fig.~\ref{SED-fit}. These very unusual and probably unrealistic examples
illustrate the difficulty to reproduce the strong spectral break present
in our objects with strong emission line galaxy spectral at low redshift.
In any case, should the near-IR photometry of our relatively bright objects be 
strongly contaminated by emission lines, these should be detectable with current
instruments.

   Two of our optical dropout galaxies (z1 and Y7) are detected by MIPS at
24\micron, leading to a preferential identification as mid-$z$ interlopers, 
and we have constraints for four additional sources (Y3, Y4, Y8
and J1, see Sect.~\ref{properties}). However, as discussed in Sect.~\ref{properties},
24 \micron\ fluxes cannot help distinguishing between high and low-$z$
solutions for a majority of our candidates. 


   In order to understand the nature of these possible 
contaminats, we have compared the present candidates/counts with the results
found our blank field survey WUDS ({\it WIRCAM Ultra Deep Survey}
\footnote{{\tt http://regaldis.ast.obs-mip.fr/}}; Pell\'o et al. in
preparation). WUDS is an extremely deep photometric survey with WIRCAM at CFHT
over $\sim$ 400{\rm arcmin}$^{-2}$ on the CFHTLS Deep pointing D3, using the same
four filter-bands $YJHKs$ as in this survey, and robust non-detection
constraints in the optical bands (i.e. $ugriz$(AB) $\sim$ 27 to 28.3 at
3$\sigma$ level, depending on filters). The main advantage of WUDS with
respect to the present survey is the large field of view and the wavelength
coverage shortwards to the $i-$band. The depth in the near-IR bands is lower,
reaching $YJ \sim$ 25.8 and $HKs \sim$ 25.3 (3$\sigma$).
When applying the same selection function introduced here for the $Y-$dropouts
(both in optical and near-IR bands), 13 candidates are retained over the WUDS
field after visual inspection, and among them 7 candidates in the $H_{AB}\sim
23.0-24.0$ interval (the same where our 5 ``bright'' candidates are
found). When we apply a more restrictive non-detection criterium in the
optical bands based the full $ugriz$ domain (detection below 2$\sigma$ in all
filters), only 3 candidates survive, all of them within the $H_{AB}\sim
23.0-24.0$ interval. Two of these WUDS candidates display the same properties as the
$z\ge7.5$ in the HAWK-I field in terms of photometric redshifts and $P(z)$
distributions, the third one beeing more dubious (degenerate solution between
low and high-$z$). This means that a more robust non-detection in the
optical bands bluewards with respect to the $I$-band
could have removed between $\sim$50 and 75\% of our present
candidates in the HAWK-I field. 



   In summary, the contamination in this field comes essentially from mid-$z$
interlopers, with a negligible contribution from late-type stars. Also strong
emission-lines seem unable to reproduce the large breaks observed. Only a
young stellar population together with a strong extinction provide a
reasonable fit at $z<4$.  Based on the comparison with the blank field survey
WUDS, and assuming that the nature of contaminant sources is the same in all
fields, we could have removed between $\sim$50 and
75\% of the present sample with a better wavelength coverage in the optical
bands bluewards from the $I$ band. 

\subsection{Comparison with previous results}
\label{comparison_others}

   We compare the number densities and properties of 
$z\ge7.5$ candidates in this sample
with those obtained by previous authors using similar
techniques to explore this redshift domain. A direct comparison is difficult
given the different selection functions. 


   Our selection criterium (c) is the same adopted by Capak et al.\ (2011)
excepted for the $Ks-4.5\mu m >0$ condition (due to partial coverage of the HAWK-I
field of view), making the comparison easier in this case. All our $Y$-dropout
candidates fulfil their color selection, except for Y4, which is formally detected in the
$z_1$ band at \ltapprox2$\sigma$ level. However, all of them are too faint to
be included in their sample (i.e. $J<$ 23.7, their 5$\sigma$ detection level), 
except for z1. This object, once corrected for magnification, 
is also $\sim$0.3 to 0.5 mags fainter than all their retained candidates
(depending on the candidate and filter). In other words, the density of bright
high-$z$ candidates in our field is consistent with the density derived by
Capak et al.\ (2011), leading to a weak constraint on the density $<3 \times
10^{-5} Mpc^{-3}$ for M$_{1500} \sim$ -23 objects. 


   The colors and SEDs of present candidates are consistent with the selection
functions introduced by Bouwens al.\ (2008, 2010) in the GOODS, HUDF, HDF South and
lensing fields.  All our candidates fulfil their $z\ge7$ preselection when
using equivalent filter-bands, i.e. our ground-based filters instead of 
$z_{850}$ and $H_{160}$ filters. Note however that even J1, which is detected
in the $z_{850}$ filter, remains in the sample due to its large break ($z_{850} -
H$ = 2.2). 
All our candidates except J1 (which is not detected in the $J-$band) fulfil
their $z\sim7$ $zJH$ selection function, as well as the selection introduced
by Hickey et al.\ (2009) for $6\ltapprox z \ltapprox 9$ galaxies. 
Instead, J1 satisfies the $z\sim9$ $JH$ selection function by Bouwens
al.\ (2008, 2010) (see also Fig.~\ref{CC_JHK}), although the detection in the
$z_{850}$ band excludes it as a genuine $z\sim9$ candidate. 
Also, two of the five candidates detected in the $Y-$band (z1 and Y3) fulfil
the rough selection defined by Ouchi et al.\ (2009) for $z\sim7$ candidates,
but they are not included when applying the selections proposed by Wilkins et
al.\ (2010) or Castellano et al.\ (2010), namely $z-Y>1.2(1.0)$ and $Y-J<2.0$. 
Our candidates are indeed slightly redder in $Y-J$, which is consistent with
the fact that all our sources have photometric redshifts $z\gtapprox7.5$. 
In summary, all the present candidates would have been selected by the usual
functions targeting $z\ge7.5$ sources based on broad-band colors. 

   Regarding the magnitudes, only two of our
candidates, namely Y8 and J1, are found in the range covered by Bouwens
al.\ (2010) in their survey of the GOODS field, i.e.\ $H_{AB}\sim 25.5-26.0$
once corrected for magnification. At this depth level, our number counts of
$z\ge7.5$ candidates are $\sim$0.06 sources {\rm arcmin}$^{-2}$, in good
agreement with their previous findings within the same magnitude interval. 

   The main difference with respect to previous studies is the presence of
several ``bright'' $M_{1500}\sim -23.0$ candidates at $z\ge7.5$ which cannot
be easily excluded based on 
broad-band colors and photometric redshifts (see also 
Sect.~\ref{contamination} above), unless a luminosity prior is used in
addition.


\section{Conclusions}
\label{conclusions}

  The photometric survey conducted on A2667 has allowed us to identify
10 $z$, $Y$ and $J$-dropout 
galaxies in the selection windows targeting z$\ge$7.5 candidates
within the $\sim 7' \times
7'$ HAWK-I field of view ($\sim$ 33{\rm arcmin}$^2$ of effective overlapping area
in all selection bands).  All of them are detected in $H$ and $Ks$ bands in
addition to $J$ and/or IRAC 3.6$\mu$m/4.5$\mu$m images, with $H_{AB}$ ranging
from 23.4 to 25.2, and modest magnification factors between 1.1 and 1.4. 
SED-fitting results in all cases yield a best solution at high-$z$
(z$\sim$7.5 to 9), with a less significant solution at low-$z$ ($z\sim$ 1.7 to
2.8). However, several of these sources seem too bright to be at $z\ge7.5$,
suggesting strong contamination by low-$z$ interlopers which must 
be also present in other current Lyman Break surveys. 

   A broad and deep wavelength coverage in the optical bands allows to suppress
the majority of low-$z$ interlopers. Indeed, based on the comparison with the
WUDS survey, we estimate that a fraction between $\sim$50-75\% of our bright
candidates could be (extreme) low-$z$ interlopers. 
The same result is achieved when photometric redshifts are computed
using a luminosity prior. In this case, only half of the sample survives, and 
only three objects (namely Y3, Y4 and Y5) are finally retained when 
including all the available information on the SED presented in this paper. 
These low-$z$ interlopers, which cannot be easily identified based on 
broad-band photometry in the optical and near-IR domains alone, 
are indeed rare objects, in the sense that they are not
well described by current spectral templates given the large break. A
reasonable good fit for these objects at $z<4$ is obtained assuming a young
stellar population together with a strong extinction. 
On the other hand, at least 1 and up to
3 sources in our sample are expected to be genuine high-$z$. Spectroscopy is needed to
ascertain their redshift and nature. Some of them could be also detected in
the IR or sub-mm bands given the estimated dust extinction and high SFRs.
Indeed, two sources in this sample, z1 and Y5, have been recently
detected in the Herschel PACS \& SPIRE bands and LABOCA, making the high-$z$
identification highly unlikely (Boone et al.\, in preparation).

   Only one source (A2667-z1) fulfils the color and magnitude selection criteria 
by Capak et al.\ (2011), although it is $\sim$0.4 magnitudes fainter than
their candidates once corrected for magnification. Its 24\micron\ flux seems
incompatible with a high-$z$ identification, although the observed flux in the
$J-$band, and the non-detection in the $I$ and $z$ bands seem difficult to
reconcile with a low-$z$ galaxy.  

After removing the brightest candidates, based on luminosity priors and SED
properties, the observed number counts of z$\gtrsim 7.5$ candidates in this
field seem to be in good agreement with expectations for an evolving LF, and also
consistent within the error bars with a constant LF since $z \sim 6$. On the
contrary, they are inconsistent with a constant LF since $z \sim 4$.

\begin{acknowledgements}
    Part of this work was supported by the French Centre National de la
Recherche Scientifique, the French Programme National de Cosmologie et
Galaxies (PNCG), as well as by the Swiss National Science
Foundation. We acknowledge support for the International Team 181 
from the International Space Science Institute in Berne.
JR acknowledges support from a EU Marie-Curie fellowship. This
work recieved support from Agence Nationale de la recherche bearing the
reference ANR-09-BLAN-0234. This paper is based on observations collected
at the European Space Observatory, Chile (71.A-0428, 082.A-0163). 
\end{acknowledgements}


\clearpage
\onecolumn
\landscape
\centering
\setlength\topmargin{0cm}
\setlength\textheight{11cm}

\endlandscape
\twocolumn

\end{document}